\newtheorem{theorem}{Theorem}
\newtheorem{proposition}{Proposition}

\newtheorem{definition}{Definition}

\newtheorem{fact}{Fact}

\newtheorem{example}{Example}

\documentclass[12pt]{article}

\usepackage{amssymb}
\usepackage{amsfonts}
\usepackage{mathrsfs}
\usepackage{graphicx}
\usepackage{amsmath}
\usepackage{subfigure}

\usepackage{tabularx}
\usepackage{booktabs}
\usepackage{caption}

\newcommand{\qed}{{\mbox{} \hspace*{\fill}{\vrule height5pt width4pt depth0pt}}\\}
\def\M{\hspace*{0.75em}}

\begin{document}
\bibliographystyle{plain}
\title{Partition-based Stability of Coalitional Games}
\author{Jian Yang\\
Department of Management Science and Information Systems\\
Business School, Rutgers University, Newark, NJ 07102}

\date{April 2023}

\maketitle
	
\begin{abstract}
	
We are concerned with the stability of a coalitional game, i.e., a transferable-utility (TU) cooperative game. First, the concept of core can be weakened so that the blocking of changes is limited to only those with multilateral backings. This principle of consensual blocking, as well as the traditional core-defining principle of unilateral blocking and one straddling in between, can all be applied to partition-allocation pairs. Each such pair is made up of a partition of the grand coalition and a corresponding allocation vector whose components are individually rational and efficient for the various constituent coalitions of the given partition. For the resulting strong, medium, and weak stability concepts, the first is core-compatible in that the traditional core exactly contains those allocations that are associated through this strong stability concept with the all-consolidated partition consisting of only the grand coalition. Probably more importantly, the latter medium and weak stability concepts are universal. By this, we mean that any game, no matter how ``poor'' it is, has its fair share of stable solutions. There is also a steepest ascent method to guide the convergence process to a mediumly stable partition-allocation pair from any starting partition. 
	
\vspace*{3mm} \noindent{\bf Keywords: }Cooperative Game; Consensual Blocking; Weak Core; Partition; Stability; Greedy Algorithm

\vspace*{.1in}\noindent{\bf JEL Code: }C71
\end{abstract}

\newpage

\section{Introduction}\label{introduction}

We are interested in providing more universally admissible stability concepts for coalitional games with transferable utility (TU). Such games are often employed to model cooperations among players to achieve their goals. Given a set $N$ of players, each game is characterized by the values $v(C)$ for various coalitions $C$ that the players form. The traditional concept of core follows the very straightforward principle of Pareto efficiency. Regrettably, only a select subset of games, namely those balanced ones, possess core members that help to divide up the gain from the grand coalition to all participants' content. This seems to go against the intuition that most if not all games should have ``natural resting states''. It might take less than that indicated in Pareto efficiency for players to not want to leave such states. 

In search of a universal stability concept, one that is applicable to all games, we naturally set our sights on partitions or coalition structures. Over the rigid grand coalition, partitions have a couple of advantages. For one thing, their evolution clearly holds a partial characterization of a set of players over a certain time horizon. Also, the ultimate outcome of such a set may just be an ordinary partition involving multiple coalitions rather than the all-consolidated partition consisting only of the grand coalition. For these reasons, partitions started to gain attention as early as in the 1970's; see, e.g., Aumann and Dreze \cite{AD1974}. 

It is tempting to follow the same principle that guides the formation of the core to develop our partition-based stability. For a given partition-allocation pair $({\cal P},\vec x)$ where ${\cal P}\equiv \{C_1,...,C_p\}$ is a partition of the grand coalition $N$ and $\vec x\equiv (x(i))_{i\in N}$ is a corresponding allocation satisfying individual rationalities of $x(i)\geq v(\{i\})$ and coalition-wise efficiencies of $\sum_{i\in C_k}x(i)=v(C_k)$, fission and fusion resistances should probably be thought of first. The former means that after any fission attempt on the current ${\cal P}$, no player could hope to gain strictly more than that is allowed by the current $\vec x$ without hurting any other player. The latter means almost the same except that ``fission'' has to be replaced by ``fusion''. %For our TU setting, this requirement is no different than that not all players could be strictly better off. This should certainly be applied to the counterpart after any fusion attempt as well. 

To attain universality, however, we have so far found it necessary to further weaken the fission-resistance notion. For a given solution $({\cal P},\vec x)$, the latter is translatable into $\sum_{i\in C'}x(i)\geq v(C')$ for every coalition $C'$ that is a strict subset of one of ${\cal P}$'s constituent coalitions. In many a real situation, this might have asked for too much. Suppose ${\cal P}$ is the all-consolidated partition $\{N\}$, for instance. The union into $N$ would often not be immediately dissolved when just one single group $C'\subsetneq N$ of players with $\sum_{i\in C'}x(i)<v(C')$ find themselves to be disgruntled. In other words, resistance to this kind of unilaterally-provoked blockings, while propping up the core concept, might have room to be weakened. 

Naturally, we think of going to the opposite end of the spectrum on blocking, that of the multilaterally-backed variety. Following the now consensual-blocking principle, an incumbent union should be allowed to dissolve only when all, rather than merely one, post-dissolution sub-coalitions could strictly profit from the process. Thus inversely, we consider a pair $({\cal P},\vec x)$ weakly fission-resistant when $\sum_{i\in C'}x(i)\geq v(C')$ for not all but just at least one $C'\notin {\cal P}$ that is constituent to any one of ${\cal P}$'s ``fission-down-to'' neighbors. When ${\cal P}=\{N\}$, weak fission resistance would lead us to allocations $\vec x$ that are in what we call the weak core. If the traditional core is capable of explaining why good marriages deserve to last, then our weak core can be used to argue why some bad marriages could still limp on.  

Weak fission resistance and the plain fusion resistance are the two axioms that define our weak stability concept. Fission resistance in its traditional core-inspired form, when coupled with the same fusion resistance, could certainly lead to a strong concept. There is a very useful middle ground as well. Here, medium fission resistance for a $({\cal P},\vec x)$-pair is just ${\cal P}$-related; it is about the worth $\sum_{C\in {\cal P}}v(C)$ dominating those same sums resulting from ${\cal P}$'s fission-down-to neighbors. Take ${\cal P}=\{N\}$ for example. When every constituent coalition of a potential fission-down-to target ${\cal P}'$ opposes the attempt, then $v(N)\geq \sum_{C\in {\cal P}'}v(C)$ for sure. When the latter is true, then any splitting attempt would meet the objection from at least one intended sub-coalition. It turns out that universality is relatively easy to verify for medium stability. This then leads to that of weak stability; see Theorem~\ref{t-monumental}. 

We can also reach a mediumly stable pair by starting from any partition and then going through a steepest ascent method (SAM). Our partitions at medium stability coincide with those identified by Apt and Radzik \cite{AR06} as $\mathbb D_{hp}$-stable. %; also, partitions at global stability correspond to their $\mathbb D_p$-stability. 
However, their stability notion is concerned with only partitions' total coalition-wise worths without spelling out allocations made to players. %We have spelled out the resistances to justify our choices on not only partitions but also allocations. 
Aside from Apt and Radzik \cite{AR06}, we are not aware of any other universality claim where all games have their own stable solutions let alone any one with our accessibility claim via SAM, that a stable solution can always be reached from any starting partition. 
%There are many variants of cooperative games besides ours. 
%Myerson \cite{M77} studied a situation where players end up in one coalition when they are linked up by a sequence of bilateral agreements. He proposed payoffs to players under each given graph formed by such links. However, untouched was the stability or supposed prominence of any particular graph, which in turn underpins an associated coalition structure. %In addition, universality was not achieved as his allocation rule is only known to meet his link-prone stability requirement when the game is superadditive. 

When players have preferences over alternatives allowed by coalitions or just over coalitions themselves, partition-based stability was achieved with strings attached; see, e.g., Greenberg and Weber \cite{GW93}, Demange \cite{D94}, Banerjee, Konishi, and Sonmez \cite{BKS01}, and Bogomolnaia and Jackson \cite{BJ02}.  
%, and Burani and Zwicker \cite{BZ03}. 
Coalition structures were considered in settings such as Ray and Vohra's \cite{RV97} involving both intra-coalition cooperation and inter-coalition competition and Yi's \cite{Y97} that couches on the partition function form (PFF). For these complex models, universality and accessibility would seem even more elusive.  %for these models, even despite the former's exclusion of fusion considerations. 
Regarding PFF, many works still revolved around the grand coalition; see, e.g., Grabisch and Funaki \cite{GF12} and Chander \cite{C14}. 

While medium stability is about the local optimality of partitions' worths, the corresponding global optimality would be the purview of the so-called optimal coalition structure. There is a huge literature devoted to the efficient computation and approximation of these optimal solutions; see, e.g., Sandholm et al. \cite{SLAST99} and Rahwan et al. \cite{RRJG09}\cite{RMWJ15}. We are also aware of the name ``weak core'' in literature. To the best of our knowledge, it usually connotes the relaxation of every $\sum_{i\in C'}x(i)\geq v(C')$ in the traditional core's definition to $\sum_{i\in C'}x(i)\geq v(C')-\epsilon$ for some $\epsilon>0$; see, e.g., Askoura \cite{A11}. Our weak concept which involves an $\forall$-to-$\exists$ switch rather than the current 0-to-$\epsilon$ relaxation.  

In the remainder, we provide model basics in Section~\ref{model}, introduce alternative core concepts in Section~\ref{weak-core}, and clarify their relationships in Section~\ref{relation}. Stability criteria based on resistances to various fission and fusion attempts are laid out in Section~\ref{stability}. We move on to Section~\ref{further} for further understandings of our stability notions. Section~\ref{pathway} is devoted to the SAM process for reaching medium stability. We conclude the paper in Section~\ref{conclusion}. 

\section{Basic Notions}\label{model}

For a set $N\equiv \{i_1,...,i_n\}$ of players for some $n\equiv |N|=1,2,...$, let $\mathscr C(N)$ be the set of $N$'s nonempty subsets which are also known as coalitions. In a coalitional game $(N,\vec v)$ with $N$ already fixed, $\vec v\equiv (v(C))_{C\in \mathscr C(N)}\in \Re^{\mathscr C(N)}$ would provide the value function and help characterize the game. Usually, there is a tacit extension to $v(\emptyset)=0$. Now let the set of allocations that are both individually rational and efficient within the game be
\begin{equation}\label{eff-allocation}
\partial(N,\vec v)\equiv \left\{\vec x\equiv (x(i))_{i\in N}\in \prod_{i\in N}[v(\{i\}),+\infty):\;\sum_{i\in N}x(i)=v(N)\right\}. 
\end{equation}
By~(\ref{eff-allocation}), $\partial(\{i\},v(\{i\}))\equiv \{v(\{i\})\}$ is nonempty for each single-player sub-game. 

Let $\mathscr P(N)$ be the space of $N$'s partitions. Each ${\cal P}\in \mathscr P(N)$ is some $\{C_1,...,C_p\}$ with

\begin{itemize}

\vspace*{-.1in}\item $C_k\in \mathscr C(N)$ for each $k=1,...,p$,

\vspace*{-.1in}\item $\bigcup_{k=1}^pC_k=N$, and

\vspace*{-.1in}\item $C_k\cap C_l=\emptyset$ for any $k\neq l$.

\end{itemize}

\vspace*{-.1in}The above could be summarized into
\begin{equation}\label{part-fact}
\sum_{C\in {\cal P}}{\bf 1}(C\ni i)=1,\hspace*{.8in}\forall i\in N,
\end{equation}
that the constituent coalitions $C$ of a partition ${\cal P}$ should cover every player $i$ exactly once. 

For a partition ${\cal P}\in\mathscr P(N)$, its set of allocations that are individually rational and also efficient within each constituent coalition $C$ for the game $(N,\vec v)$ is 
\begin{equation}\label{allocation-def}\begin{array}{l}
\mathscr X(N,\vec v,{\cal P})\equiv \prod_{C\in {\cal P}}\partial(C,\vec v|_{\mathscr C(C)})\\
\;\;\;\;\;\;\;\;\;\;\;\;\;\;\;\;\;\;\;=\left\{\vec x\equiv (x(i))_{i\in N}\in \prod_{i\in N}[v(\{i\}),+\infty):\;\sum_{i\in C}x(i)=v(C),\;\;\;\forall C\in {\cal P}\right\}.
\end{array}\end{equation}
By~(\ref{eff-allocation}) and~(\ref{allocation-def}), $\mathscr X(N,\vec v,\{N\})=\partial(N,\vec v)$. According to~(\ref{allocation-def}), each coalition $C$ within ${\cal P}$ would see its value $v(C)$ exactly divided among players $i$ in the coalition via $\vec x$'s to-$C$ projection $\vec x|_{C}\equiv (x(i))_{i\in C}$. Let $\underline{\cal P}\equiv \{\{i_1\},...,\{i_n\}\}$ be the all-splintered partition. Following the comment after~(\ref{eff-allocation}), at least $\mathscr X(N,\vec v,\underline{\cal P})$ is nonempty with the only member $\underline{\vec x}\equiv (v(\{i\}))_{i\in N}$. %In view of~(\ref{eff-allocation}) and~(\ref{allocation-def}), $\partial(N,\vec v)=\mathscr X(N,\vec v,\{N\})$.  

We let the set of feasible partition-allocation pairs be 
\begin{equation}\label{pair-def}
\mathscr Q(N,\vec v)\equiv \bigcup_{{\cal P}\in \mathscr P(N)}\{{\cal P}\}\times \mathscr X(N,\vec v,{\cal P}).
\end{equation}
Note $\mathscr Q(N,\vec v)$ of~(\ref{pair-def}) is nonempty as it contains at least $(\underline{\cal P},\underline{\vec x})$. We want to establish stability notions $\mathbb S$ so that for any $\vec v\in \Re^{\mathscr C(N)}$, not necessarily a privileged one say that which is balanced, each such $\mathbb S$ would provide a nonempty subset $\mathbb S(N,\vec v)$ of $\mathscr Q(N,\vec v)$. That is, we strive for ``universality'' in the sense that 
\begin{equation}\label{universal-def}
\mathbb S(N,\vec v)\neq\emptyset\;\;\;\mbox{ or equivalently }\;\;\;|\mathbb S(N,\vec v)|\geq 1,\hspace*{.8in}\forall \vec v\in \Re^{\mathscr C(N)}.
\end{equation}
On the flip side, we also stress ``reasonableness''. The notions should meet certain expectations; better still, they should be compatible with some existing notions when the game is ``good'' enough. An  $\mathbb S'$ would be tighter than $\mathbb S$ when $\mathbb S'(N,\vec v)\subseteq \mathbb S(N,\vec v)$ for every $\vec v \in \Re^{\mathscr C(N)}$.

The existing notion of core might have stressed too much on the grand coalition $N$ or correspondingly, the all-consolidated partition $\{N\}$; even when the focus is there, it might have left room for further loosening. For a game $(N,\vec v)$, its traditional core is 
\begin{equation}\label{core-def}
\mathbb X^+(N,\vec v)\equiv \left\{\vec x\in\partial(N,\vec v):\;\sum_{i\in C}x(i)\geq v(C),\;\;\;\forall C\in \mathscr C(N)\hspace*{-.02in}\setminus\hspace*{-.03in}\{N\}\right\}.
\end{equation}
By~(\ref{eff-allocation}) and~(\ref{allocation-def}), the requirement that each $x(i)\geq v(\{i\})$ imposed by~(\ref{core-def}) is certainly redundant. Here as well as similar places later, we allow such redundancies out of convenience. 

By Bondareva \cite{B63} and Shapley \cite{S67}, $\mathbb X^+(N,\vec v)$ of~(\ref{core-def}) would be nonempty if and only if $(N,\vec v)$ is balanced in the sense that $v(N)$ is above 
\begin{equation}\label{balance-lp}\begin{array}{lll}
z^+(N,\vec v)\equiv &\max& \sum_{C\in\mathscr C(N)}v(C)\cdot \delta(C)\\
& \mbox{ s.t. } & \sum_{C\in\mathscr C(N)\mbox{ and }C\ni i}\delta(C)=1\hspace*{.8in}\forall i\in N\\
& & \delta(C)\in \Re_+\equiv[0,+\infty)\hspace*{.8in}\forall C\in\mathscr C(N).
\end{array}\end{equation}
Certainly, only a select few of $\vec v$'s have $v(N)\geq z^+(N,\vec v)$ of~(\ref{balance-lp}). 

\section{Other Core Concepts}\label{weak-core}

Suppose Alice wants to end an unhappy marriage with Bob who does not contribute much to the family's well being including the raising of their daughter Carol. Thus, the dilemma faced by her is merely between staying put and taking Carol to leave Bob. 
	
In the spirit of~(\ref{core-def}) which condones any unilateral change of the status quo, Alice's attempt would succeed as long as she and Carol could fare better by staying away from Bob. %, without caring any bit about the change to his welfare. Reality, on the other hand, might be more complex. Very likely, 
In reality, Bob might try hard to prevent the two from leaving him unless his fear of precipitous falls in his living standards is assuaged. In any event, such a breakaway would be more realizable when the intended change becomes consensual, like when Alice and Carol could still be better off after Bob is compensated enough for him to continue on with his life style. 

More concretely, suppose Alice or Carol alone would get 0 because they need each other; also, suppose Bob and Carol just by them two would get 0 because Bob is incapable of taking care of Carol; in addition, Alice and Bob by them two would get 0. 

Now, suppose Alice gets 0, Bob 6, and Carol 2 within the marriage; Alice and Carol by them two can get 6; also, Bob alone gets 4. Then, the traditional theory would deem their marriage unstable simply because a brighter alternative calls for the Alice-Carol grouping. However, since Bob thinks he is getting a raw deal out of the potential divorce, he may fight tooth and nail to thwart Alice's attempt to leave him. 

One way for Alice to quickly exit with Carol might be for her to commit a transfer of 3 to Bob once after the dissolution of the union. Thus, for the $(A=0,B=6,C=2)$-unhappy marriage to crumble, it would take an outside alternative of $(AC''=6-3,B''=4+3)=(3,7)$ with both $A+C=0+2<3=AC''$ and $B=6<7=B''$. 

In contrast, instability in the old way of $(A=0,B=6,C=2)$ in front of $(AC'=6,B'=4)$ would only take $A+C=0+2<6=AC'$. When stated oppositely, stability in the old way stipulates the blocking of all unilateral changes of the status quo. It is formulated in $\mathbb X^+(N,\vec v)$ of~(\ref{core-def}), which could be rewritten as
\begin{equation}\label{core-def-trad}
\left\{\vec x\in \partial(N,\vec v):\;\forall C'\in {\cal P}'\mbox{ we have }\sum_{i\in C'}x(i)\geq v(C'),\hspace*{.1in}\forall {\cal P}'\in \mathscr P(N)\hspace*{-.02in}\setminus\hspace*{-.03in}\{\{N\}\}\right\}.
\end{equation}
By~(\ref{core-def-trad}), it takes $\vec x|_{C'}\equiv (x(i))_{i\in C'}$'s resistance against $v(C')$ at all constituent coalitions $C'$ of any non-$\{N\}$ partition, for the payoff vector $\vec x$ to be stable in the traditional sense. 

The blocking advocated by~(\ref{core-def-trad}) is wider than that of merely consensual changes as reflected in the Alice-Bob-Carol (ABC) story. Let us formalize the latter in the weak core $\mathbb X^-(N,\vec v)$: 
\begin{equation}\label{core-def-weak}
\left\{\vec x\in \partial(N,\vec v):\;\exists C'\in {\cal P}'\mbox{ so that }\sum_{i\in C'}x(i)\geq v(C'),\hspace*{.1in}\forall {\cal P}'\in \mathscr P(N)\hspace*{-.02in}\setminus\hspace*{-.03in}\{\{N\}\}\right\}.
\end{equation}
By~(\ref{core-def-weak}), it would take the same resistance at merely one representative coalition $C'$ of each non-$\{N\}$ partition, for $\vec x$ to be stable in our weak sense. 

For convenience, let us call the traditional $\mathbb X^+(N,\vec v)$ the strong core. In some sense, it gives any one constituent coalition of any non-$\{N\}$ partition some outsize veto power on the potential fusion of the partition into $\{N\}$; quite oppositely, the weak core $\mathbb X^-(N,\vec v)$ gives the same coalition some outsize veto power on the potential fission of $\{N\}$ into the said non-$\{N\}$ partition. While the latter still fits the usual stability meaning of ``a state that is difficult to leave once there'', both might have erred on giving too much power to one single coalition. 

There, though, is a potential compromise. In the ABC story, we deem $(A=0,B=6,C=2)$-union dissolvable in the weak sense only after reaching the post-transfer one of $(AC''=3,B''=7)$. However, $A+B+C=0+6+2=8<10=6+4=AC'+B'$ is what really matters here. In the presence of this strict inequality, Alice could always work out something to make Bob happy to leave on the one hand and on the other hand, to still leave her and Carol better off than in the marriage. Thus, the midway concept should render the $(A=0,B=6,C=2)$-union already dissolvable into the $(AC'=6,B'=4)$-one. 

Formally, let us define the worth of a partition ${\cal P}\in \mathscr P(N)$ in a game $\vec v\in \Re^{\mathscr C(N)}$ by
\begin{equation}\label{worth-def}
\tilde w(N,\vec v,{\cal P})\equiv \sum_{C\in {\cal P}}v(C).
\end{equation} 
Certainly, $\tilde w(N,\vec v,\{N\})=v(N)$ and $\tilde w(N,\vec v,\underline{\cal P})=\sum_{i\in N}v(\{i\})$. By~(\ref{eff-allocation}),~(\ref{allocation-def}), and~(\ref{worth-def}), 
\begin{equation}\label{youxiao}
\mathscr X(N,\vec v,\{N\})=\partial(N,\vec v)\neq\emptyset\;\;\;\Longleftrightarrow\;\;\;v(N)\geq \tilde w(N,\vec v,\underline{\cal P}).
\end{equation}

We then define the medium core $\mathbb X^0(N,\vec v)$ as
\begin{equation}\label{core-def-medium}
\left\{\begin{array}{ll}
\mathscr \partial(N,\vec v) & \;\;\;\;\;\;\mbox{ when }v(N)\geq \max_{{\cal P}\in \mathscr P(N)\setminus\{N\}}\tilde w(N,\vec v,{\cal P}),\\
\emptyset & \;\;\;\;\;\;\mbox{ otherwise}.
\end{array}\right.\end{equation}
Since $v(N)\geq \max_{{\cal P}\in \mathscr P(N)\setminus\{N\}}\tilde w(N,\vec v,{\cal P})$ supersedes $v(N)\geq \tilde w(N,\vec v,\underline{\cal P})$, the set $\partial(N,\vec v)$ in the first line of~(\ref{core-def-medium}) must be nonempty. As argued, all that takes to break the grand coalition $N$ in the current medium sense is one non-$\{N\}$ partition ${\cal P}$ whose worth $\tilde w(N,\vec v,{\cal P})$ is strictly above $v(N)$. We leave out the details on how the constituent coalitions $C$ of ${\cal P}$ could work out mutual compensations to incentivize the transition from $\{N\}$ to ${\cal P}$. 

\section{Natural Relationships}\label{relation} 

The definitions~(\ref{core-def-trad}) to~(\ref{core-def-medium})  could lead to not only 
\begin{equation}\label{weaker-core}
\mathbb X^+(N,\vec v)\subsetneq \mathbb X^0(N,\vec v)\subsetneq X^-(N,\vec v),
\end{equation}
but also for the second inclusion relationship, the more detailed property that
\begin{equation}\label{detail}
v(N)\geq \max_{{\cal P}\in \mathscr P(N)\setminus\{N\}}\tilde w(N,\vec v,{\cal P})\;\;\;\Longrightarrow\;\;\;\mathbb X^-(N,\vec v)=X^0(N,\vec v)=\partial(N,\vec v).
\end{equation}

The first inclusion relationship in~(\ref{weaker-core}) can be summarized in the following.

\begin{proposition}\label{strong-plain}
Between $\mathbb X^+(N,\vec v)$ of~(\ref{core-def-trad}) and $\mathbb X^0(N,\vec v)$ of~(\ref{core-def-medium}),
\[ \mathbb X^+(N,\vec v)\subseteq \mathbb X^0(N,\vec v),\hspace*{.8in}\forall \vec v\in \Re^{\mathscr C(N)}.\]
\end{proposition}

\noindent{\bf Proof of Proposition~\ref{strong-plain}: }Consider the following linear program:
\begin{equation}\label{p-prime}\begin{array}{lll}
z^0(N,\vec v)\equiv & \min & w\\
& \mbox{s.t. } & w\geq \tilde w(N,\vec v,{\cal P})\hspace*{.8in}\forall {\cal P}\in \mathscr P(N)\\
& & w\in \Re.
\end{array}\end{equation}
By~(\ref{core-def-medium}) and~(\ref{p-prime}),
\begin{equation}\label{yeman}
v(N)\geq z^0(N,\vec v)\;\;\;\Longleftrightarrow\;\;\;\mathbb X^0(N,\vec v)=\partial (N,\vec v)\neq\emptyset.
\end{equation} 

Meanwhile,~(\ref{p-prime})'s dual is 
\begin{equation}\label{d-prime}\begin{array}{lll}
z^0(N,\vec v)\equiv & \max & \sum_{{\cal P}\in \mathscr P(N)}\tilde w(N,\vec v,{\cal P})\cdot \lambda({\cal P})\\
& \mbox{s.t. } & \sum_{{\cal P}\in \mathscr P(N)}\lambda({\cal P})=1\\
& & \lambda({\cal P})\in \Re_+\hspace*{.8in}\forall {\cal P}\in \mathscr P(N).
\end{array}
\end{equation}
A feasible solution for~(\ref{d-prime}) would provide one for~(\ref{balance-lp}): letting $\delta(C)=\sum_{{\cal P}\ni C}\lambda({\cal P})$ for any $C\in \mathscr C(N)$ while noting~(\ref{part-fact}), we could achieve for any $i\in N$ that
\begin{equation}\label{more-f}
1=\sum_{{\cal P}\in \mathscr P(N)}\lambda({\cal P})\cdot \sum_{C\in {\cal P}}{\bf 1}(C\ni i)=\sum_{C\in \mathscr C(N)}{\bf 1}(C\ni i)\cdot\sum_{{\cal P}\ni C}\lambda({\cal P})=\sum_{C\in \mathscr C(N)\mbox{ and }C\ni i}\delta(C).
%\vec 1=\sum_{{\cal P}\in \mathscr P(N)}\lambda({\cal P})\cdot \sum_{C\in {\cal P}}\vec\chi(C)=\sum_{C\in \mathscr C(N)}\vec\chi(C)\cdot\sum_{{\cal P}\ni C}\lambda({\cal P})=\sum_{C\in \mathscr C(N)}\delta(C)\cdot\vec\chi(C).
\end{equation}

With~(\ref{more-f}), we have $z^0(N,\vec v)\leq z^+(N,\vec v)$ of~(\ref{balance-lp}). Therefore,
\begin{equation}\label{last-ineq}
\mathbb X^+(N,\vec v)\neq\emptyset\;\;\;\Longleftrightarrow\;\;\; v(N)\geq z^+(N,\vec v)\;\;\;\Longrightarrow\;\;\; v(N)\geq z^0(N,\vec v).
\end{equation}
By~(\ref{core-def-trad}), $\mathbb X^+(N,\vec v)$ would be contained in $\partial(N,\vec v)$ even when it is nonempty. Thus,~(\ref{yeman}) and~(\ref{last-ineq}) would together entail the desired inclusion result.\qed

The converse to the first inclusion relationship of~(\ref{weaker-core}) is not true. We have a counter example where $\mathbb X^+(N,\vec v)=\emptyset$ and yet $\mathbb X^0(N,\vec v)\neq\emptyset$. 

\begin{example}\label{example-a}Consider $N=\{1,2,3\}$, $v(\{1\})=v(\{2\})=v(\{3\})=0$, $v(\{1,2\})=v(\{1,3\})=v(\{2,3\})=5$, and $v(\{1,2,3\})=6$. Here, $\{N\}$ is dominant in its worth of 6 over others which top at 5; yet, there is no core member as $x(1)+x(2)+x(3)=6$ contradicts with $x(1)+x(2)\geq 5$, $x(1)+x(3)\geq 5$, and $x(2)+x(3)\geq 5$.  
\end{example}

Since $\mathbb X^-(N,\vec v)$ by its definition at~(\ref{core-def-weak}) is contained in $\partial(N,\vec v)$, to verify the more detailed expression~(\ref{detail}) for the second inclusion in~(\ref{weaker-core}), we need just the following. 

\begin{proposition}\label{p-wb}
For $\mathbb X^-(N,\vec v)$ of~(\ref{core-def-weak}),
\[ v(N)\geq \max_{{\cal P}\in \mathscr P(N)\setminus\{N\}}\tilde w(N,\vec v,{\cal P})\;\;\;\Longrightarrow\;\;\;\partial(N,\vec v)\subseteq \mathbb X^-(N,\vec v),\hspace*{.8in}\forall \vec v\in\Re^{\mathscr C(N)}.\]
\end{proposition}	

\noindent{\bf Proof of Proposition~\ref{p-wb}: }Suppose $v(N)\geq \max_{{\cal P}\in \mathscr P(N)\setminus\{N\}}\tilde w(N,\vec v,{\cal P})$. By the comment after~(\ref{core-def-medium}), we know $\partial(N,\vec v)$ is nonempty. Let $\vec x\in \partial(N,\vec v)$. By~(\ref{eff-allocation}),~(\ref{worth-def}), and the hypothesis, 
\begin{equation}\label{bath}
\sum_{C\in {\cal P}}\sum_{i\in C}x(i)=\sum_{i\in N}x(i)=v(N)\geq \tilde w(N,\vec v,{\cal P})=\sum_{C\in {\cal P}}v(C),\hspace*{.5in}\forall {\cal P}\in \mathscr P(N)\hspace*{-.02in}\setminus\hspace*{-.03in}\{N\}.
\end{equation} 

For any ${\cal P}\in \mathscr P(N)\hspace*{-.02in}\setminus\hspace*{-.03in}\{N\}$,~(\ref{bath}) must lead to  the existence of one $C\in {\cal P}$ such that $\sum_{i\in C}x(i)\geq v(C)$. In view of~(\ref{core-def-weak}), this just means that $\vec x\in \mathbb X^-(N,\vec v)$. \qed

The converse to the second inclusion relationship of~(\ref{weaker-core}) is again not true. The ABC story has provided a counter example where $\mathbb X^0(N,\vec v)=\emptyset$ and yet $\mathbb X^-(N,\vec v)\neq\emptyset$. 

%\begin{example}\label{example-b}
%For the case where $N=\{1,2\}$, $v(\{1\})=v(\{2\})=4$, and $v(\{1,2\})=6$, note $\vec x=(0,6)$ is in the weak core. While player 1 wants to leave the union of the two players, player 2 would block her attempt. Yet, $z(N,\vec v)=v(\{1\})+v(\{2\})=8>6=v(N)$. 
%\end{example}

\begin{example}\label{example-b}
Consider the case where $N=\{A,B,C\}$, $v(\{A\})=0$, $v(\{B\})=4$, $v(\{C\})=0$, $v(\{A,B\})=0$, $v(\{A,C\})=6$, $v(\{B,C\})=0$, and $v(\{A,B,C\})=8$. 

Note $\vec x=(0,6,2)$ is in the weak core. Under this allocation plan, no attempt to split the grand coalition $N$ can make all resulting sub-coalitions strictly better off. The most competitive alternative comes from the partition $\{\{A,C\},\{B\}\}$. While the sub-coalition $\{A,C\}$ has strong incentives to leave the grand coalition, player $B$ would hold out. %Player 2 also wants to leave $\{1,3\}$ behind, but the two players involved in the latter would not agree. The same can be said about player 3. 

On the other hand, 
\[ z^0(N,\vec v)=\tilde w(N,\vec v,\{\{A,C\},\{B\}\})=6+4=10>8=\tilde w(N,\vec v,\{N\})=v(N).\] 
Thus, the grand coalition does not attain the highest worth here. 
\end{example}

When players in Example~\ref{example-b} start from the partition $\{\{A,C\},\{B\}\}$, they would definitely not want to form the grand coalition $N$ with a lower worth. However, this should be no reason to refute our weak notion because by hosting states ``that are difficult to leave once there'', the concept nevertheless holds the essence of stability. On the flip side, how players ``reach'' the state of $\{N\}=\{\{A,B,C\}\}$ along with the allocation $\vec x=(0,6,2)$ should be of less a concern. This combination is stable enough in our view simply because once reached, any attempt to change the status quo would be reasonably thwarted. 

The guiding philosophy for the traditional core, on the other hand, appears to be the more optimality-themed one of ``being difficult to reach when not yet there''. Just the widely adopted term ``blocking'' has already the connotation of ``an obstacle on the way towards the formation''. After all, a core member is one that overcomes all such obstacles including the very unilateral ones that are initiated by lone sub-coalitions. In their tolerant attitudes towards self-centered sub-coalitions,~(\ref{core-def-trad}) and~(\ref{core-def-weak}) are rather alike than apart. It is only that~(\ref{core-def-trad}) allows any single one already-satisfied and hence only-to-lose sub-coalition to prevent any further fusion into $\{N\}$; meanwhile,~(\ref{core-def-weak}) allows any single one already-satisfied and hence only-to-lose sub-coalition to prevent any further fission from $\{N\}$. 

%Suppose Alice and Bob can each individually earn 4, while the total gain from their marriage is 6. Then, when starting as singles, they would not enter into the marriage because there is no way to make both happier. However, when starting from a (1,5)-allocation in the marriage, Alice's attempt to break away from the union would receive the objection from Bob as vehemently as when he would defend his bachelorship in the alternative situation. 

%The above being said, we do have ``pickier'' and probably to some audience more reasonable stability notions that are not only still universal but also capable of shunning away from such a ``bad'' choice as $(\{\{1,2,3\}\},(3,-1,-1))$. Moreover, we may note from  Example~\ref{example-b} that while every $v(C)>0$, a weak core members $\vec x$ could still have $x(i)\leq 0$. This is not possible for the traditional core simply because of the individual-rationality requirements. 

Finally, when it comes to the medium core of~(\ref{core-def-medium}), there could be an alternative expression after consulting~(\ref{eff-allocation}),~(\ref{worth-def}), and~(\ref{youxiao}):
\begin{equation}\label{core-def-middle}
\mathbb X^0(N,\vec v)\equiv \left\{\vec x\in \partial(N,\vec v):\;\sum_{C'\in {\cal P}'}\sum_{i\in C'}x(i)\geq \sum_{C'\in {\cal P}'}v(C'),\hspace*{.1in}\forall {\cal P}'\in \mathscr P(N)\hspace*{-.02in}\setminus\hspace*{-.03in}\{\{N\}\}\right\}.
\end{equation}
The style of~(\ref{core-def-middle}) is actually more consistent with those of~(\ref{core-def-trad}) and~(\ref{core-def-weak}). From it the inclusion relationships in~(\ref{weaker-core}) would probably be more obvious. 

\section{General Stability Notions}\label{stability}

As mentioned, we aim at stability notions that are subsets of $\mathscr Q(N,\vec v)$ defined at~(\ref{pair-def}). The various core concepts are merely ones coupled with the all-consolidated partition $\{N\}$. On the other hand, they provide general principles that lead us to the more general notions. 

Let us adopt the usual refining/coarsening partial order between some pairs of partitions ${\cal P}\equiv \{C_1,...,C_p\}$ and ${\cal P}'\equiv \{C'_1,...,C'_{p'}\}$, with the orientation favoring consolidation. We write ${\cal P}\prec{\cal P}'$ when $p\geq p'+1$ and each $C'_{k'}$ is the union of some $C_k$'s. In words, we can reach ${\cal P}$ from ${\cal P}'$ via splits or fission-down steps and reach ${\cal P}'$ from ${\cal P}$ via mergers or fusion-up steps. Note a split could involve multiple constituent coalitions and each such coalition could be split into more than two sub-coalitions. Only when one two-way split is involved, would a split result in $p=p'+1$. A similar observation can be made about a merger. 

Given any ${\cal P}\in\mathscr P(N)$, we define its fission-down-to neighborhood as 
\begin{equation}\label{fi-def}
\mathscr I(N,{\cal P})\equiv \{{\cal P}'\in\mathscr P(N):\;{\cal P}'\prec{\cal P}\}.
\end{equation}
Similarly, we define its fusion-up-to neighborhood as
\begin{equation}\label{fu-def}
\mathscr U(N,{\cal P})\equiv \{{\cal P}'\in\mathscr P(N):\;{\cal P}\prec{\cal P}'\}.
\end{equation}
Clearly, $\mathscr I(N,\{N\})=\mathscr P(N)\hspace*{-.02in}\setminus\hspace*{-.03in}\{\{N\}\}$, $\mathscr U(N,\{N\})=\emptyset$ as the all-consolidated partition has no room for further fusion, $\mathscr I(N,\underline{\cal P})=\emptyset$ as the all-splintered partition $\underline{\cal P}\equiv\{\{i_1\},...,\{i_n\}\}$ has no room for further fission, and $\mathscr U(N,\underline{\cal P})=\mathscr P(N)\hspace*{-.02in}\setminus\hspace*{-.03in}\{\underline{\cal P}\}$.

Under a given value function $\vec v\in\Re^{\mathscr C(N)}$, the various core definitions at~(\ref{core-def-trad}),~(\ref{core-def-weak}), and~(\ref{core-def-middle}) naturally inspire the following potential criteria, in the strong-medium-weak order. 

\begin{definition}\label{def-fi-strong}
A partition-allocation pair $({\cal P},\vec x)\in \mathscr Q(N,\vec v)$ of~(\ref{pair-def}) is strongly fission-resistant in the face of $(N,\vec v)$ when for any ${\cal P}'\in \mathscr I(N,{\cal P})$ of~(\ref{fi-def}),  
\[ \forall C'\in {\cal P}'\hspace*{-.02in}\setminus\hspace*{-.03in}{\cal P}\;\;\;\mbox{ we have }\;\;\;\sum_{i\in C'}x(i)\geq v(C'),\]
where by~(\ref{allocation-def}), the right-hand side is also $\sum_{i\in C'}x'(i)$ for any $\vec x^{\,\prime}\in\mathscr X(N,\vec v,{\cal P}')$. 
\end{definition}

\begin{definition}\label{def-fi-medium}
A partition-allocation pair $({\cal P},\vec x)\in \mathscr Q(N,\vec v)$ of~(\ref{pair-def}) is mediumly fission-resistant in the face of $(N,\vec v)$ when for any ${\cal P}'\in \mathscr I(N,{\cal P})$ of~(\ref{fi-def}),  
\[ \sum_{i\in N}x(i)=\sum_{C\in {\cal P}}\sum_{i\in C}x(i)=\sum_{C\in {\cal P}}v(C)=\tilde w(N,\vec v,{\cal P})\geq \tilde w(N,\vec v,{\cal P}'),\]
where the equalities are due to~(\ref{part-fact}),~(\ref{allocation-def}), and~(\ref{worth-def}). 
\end{definition}

\begin{definition}\label{def-fi-weak}
A partition-allocation pair $({\cal P},\vec x)\in \mathscr Q(N,\vec v)$ of~(\ref{pair-def}) is weakly fission-resistant in the face of $(N,\vec v)$ when for any ${\cal P}'\in \mathscr I(N,{\cal P})$ of~(\ref{fi-def}),  
\[ \exists C'\in {\cal P}'\hspace*{-.02in}\setminus\hspace*{-.03in}{\cal P}\;\;\;\mbox{ so that }\;\;\;\sum_{i\in C'}x(i)\geq v(C'),\]
where by~(\ref{allocation-def}), the right-hand side is also $\sum_{i\in C'}x'(i)$ for any $\vec x^{\,\prime}\in\mathscr X(N,\vec v,{\cal P}')$. 
\end{definition}

By Definition~\ref{def-fi-strong}, any attempt to further split a strongly fission-resistant pair $({\cal P},\vec x)$ would be objected by all affected sub-coalitions $C'$. This is consistent with the message of~(\ref{core-def-trad}) for the special case with ${\cal P}=\{N\}$. By Definition~\ref{def-fi-medium}, the total payoff associated with a mediumly fission-resistant pair $({\cal P},\vec x)$ is high enough to give players certain wherewithals to resist further splittings. This is consistent with the message of~(\ref{core-def-middle}) for ${\cal P}=\{N\}$. Judging from its final inequality, this definition is ultimately about ${\cal P}$ alone, so long as $\vec x$'s membership to $\mathscr X(N,\vec v,{\cal P})$ has been ensured. 

By Definition~\ref{def-fi-weak}, any attempt to further split a weakly fission-resistant pair $({\cal P},\vec x)$ should be objected by at least one affected sub-coalition $C'$ with vested interests in maintaining the status quo. Note the requirement that $C'\notin {\cal P}$ ensures that the $C'$ in existence is a sub-coalition in the strict sense. When ${\cal P}$ happens to be $\{N\}$, a constituent coalition $C'$ of any partition ${\cal P}'$ in the fission-down-to neighborhood $\mathscr I(N,{\cal P}=\{N\})$ would not be $N$ itself. Thus, the requirement $C'\in {\cal P}'\hspace*{-.02in}\setminus\hspace*{-.03in}\{N\}$ only amounts to $C'\in {\cal P}'$. This may help us see that~(\ref{core-def-weak}) is indeed consistent with this definition.  

For a non-$\{N\}$ partition ${\cal P}$ to be stable, it has to resist temptations from fission-down-to neighbors but also fusion-up-to ones. The following notion is thus natural. 

\begin{definition}\label{def-fu}
A partition-allocation pair $({\cal P},\vec x)\in \mathscr Q(N,\vec v)$ of~(\ref{pair-def}) is fusion-resistant in the face of $(N,\vec v)$ when for any ${\cal P}'\in \mathscr U(N,{\cal P})$ of~(\ref{fu-def}) and $C'\in {\cal P}'\hspace*{-.02in}\setminus\hspace*{-.03in}{\cal P}$,
\[ \sum_{i\in C'}x(i)=\sum_{C\in {\cal P},\,C\subsetneq C'}\;\sum_{i\in C}x(i)=\sum_{C\in {\cal P},\,C\subsetneq C'}v(C)\geq v(C'),\]
where the equalities are due to~(\ref{allocation-def}) and~(\ref{pair-def}). 
\end{definition}

By Definition~\ref{def-fu}, a fusion-resistant pair $({\cal P},\vec x)$ should have no strict incentive to merge further, for no player in any super-coalition $C'$ made up of existing coalitions could strictly benefit from it without hurting anyone else. Judging from the final inequality, it is ultimately a property of ${\cal P}$ alone, so long as $\vec x$'s membership to $\mathscr X(N,\vec v,{\cal P})$ has been given. %The same requirement here simply makes the definition more symmetric-looking. When $C'\in {\cal P}$, we would have $\sum_{i\in C'}x(i)=v(C')$ as $({\cal P},\vec x)\in \mathscr Q(N,\vec v)$ of~(\ref{pair-def}) and hence $\vec x\in \mathscr X(N,\vec v,{\cal P})$ of~(\ref{allocation-def}).   

The split of an existing coalition affects all intended sub-coalitions while the merger of some existing coalitions does not, at least in our current coalition- rather than partition-form, affect other coalitions outside the targeted super-coalition. %It is perfect alright for the holdout of just one small clique within an existing coalition to prevent any further split and at the same time, for any extra benefit to all members involved in a group of coalitions to muster the group's merger.  
This might explain the discrepancy in the numbers of varieties that exist between fission and fusion resistances. Concerning the former, different attitudes toward ``collateral damages'' can already rationalize the schism between~(\ref{core-def-trad}) and~(\ref{core-def-weak}) with~(\ref{core-def-middle}) caught in the middle. For the latter, however, there is no collateral damage attached to any unrelated party to a merger.  

Now, let all feasible partition-allocation pairs $({\cal P},\vec x)$ in $\mathscr Q(N,\vec v)$ of~(\ref{pair-def}) that satisfy Definition~\ref{def-fi-strong} form the set $\mathbb Q^{\mbox{i}+}(N,\vec v)$, those that satisfy Definition~\ref{def-fi-medium} the set $\mathbb Q^{\mbox{i}0}(N,\vec v)$, and those that satisfy Definition~\ref{def-fi-weak} the set $\mathbb Q^{\mbox{i}-}(N,\vec v)$; also, let all feasible partition-allocation pairs that satisfy Definition~\ref{def-fu} form the set $\mathbb Q^{\mbox u}(N,\vec v)$. The stabilities of interest are 
\begin{equation}\label{stab-def}
\mathbb S^*(N,\vec v)\equiv \mathbb Q^{\mbox{i}\ast}(N,\vec v)\cap \mathbb Q^{\mbox u}(N,\vec v),\hspace*{.8in}\forall \vec v\in \Re^{\mathscr C(N)},
\end{equation}
where $\ast$ could be $+$, $0$, or $-$. They are all clearly axiom-driven. %, with Definitions~\ref{def-fi-strong} to~\ref{def-fu} serving as the defining axioms. 

The rationales for the core of~(\ref{core-def-trad}) can be used to justify $\mathbb S^+$ because it is ``strong-core-compatible''. Rigorously, let us understand this already-adopted term in the sense that any $(\{N\},\vec x)$ would be strongly stable if and only if $\vec x$ is a member of the strong core. Similar observations can be made for the medium and weak notions. While the given solution pair is automatically fusion-resistant by Definition~\ref{def-fu} as $\mathscr U(N,\{N\})=\emptyset$, it is clear from~(\ref{core-def-trad}),~(\ref{core-def-weak}),~(\ref{core-def-middle}), and~(\ref{stab-def}), as well as Definitions~\ref{def-fi-strong} to~\ref{def-fi-weak} that  
\begin{equation}\label{badaling}\begin{array}{l}
\mathbb S^*(N,\vec v)\cap \left[\{\{N\}\}\times \partial(N,\vec v)\right]=\mathbb Q^{\mbox{i}\ast}(N,\vec v)\cap \left[\{\{N\}\}\times \partial(N,\vec v)\right]\\
\;\;\;\;\;\;\;\;\;\;\;\;\;\;\;\;\;\;\;\;\;\;\;\;\;\;\;\;\;\;\;\;\;\;\;\;\;\;\;\;\;\;\;\;\;\;=\{\{N\}\}\times \mathbb X^\ast(N,\vec v).
\end{array}\end{equation}
%where $\ast$ could be $+$, $0$, or $-$. 

For these stability notions, we shall establish that  
\begin{equation}\label{3-way}
\mathbb S^+(N,\vec v)\subseteq \mathbb S^0(N,\vec v)\subseteq \mathbb S^-(N,\vec v),\hspace*{.8in}\forall \vec v\in \Re^{\mathscr C(N)};
\end{equation}
\begin{equation}\label{universal-studio}
\mathbb S^0(N,\vec v)\neq\emptyset,\hspace*{.8in}\forall \vec v\in\Re^{\mathscr C(N)}.
\end{equation}
With~(\ref{3-way}) and~(\ref{universal-studio}), we know that not only medium stability $\mathbb S^0$ is universal, but also so is the indeed looser weak stability $\mathbb S^-$. 

As for the indeed tighter strong stability $\mathbb S^+$, Example~\ref{example-a} could also serve as a counter example to it being universal. In it, none of the non-$\{N\}$ partitions is fusion-resistant. But as shown, no $N$-efficient and individually rational allocation $\vec x$ could make the pair $(\{N\},\vec x)$ strongly fission-resistant. On the other hand, many $\vec x$'s could render $(\{N\},\vec x)$ midiumly or weakly fission-resistant. Take the example of $\vec x=(2,2,2)$. Even though players 1 and 2 may agitate for their own $\{1,2\}$-coalition, the very content player 3 would thwart their attempt. 

Even still non-universal, the general-${\cal P}$-based notion $\mathbb S^+$ has already its advantage over the traditional $\{N\}$-centric core notion. When faced with any $\vec v\in\Re^{\mathscr C(N)}$, we believe ``real players'' will eventually have to ``settle down at'' some solution pair $({\cal P},\vec x)$. If this were true, then $\mathbb S^+$ would have more chance of emulating reality than the core. When $N=\{1,2\}$ and $v(\{1\})=v(\{2\})=v(\{1,2\})=1$, for instance, $\mathbb S^+$ would contain $(\{\{1\},\{2\}\},(1,1))$ as a naturally stable point; whereas, insisting on the core would leave us empty-handed.        

Of course, it would be even better if $\mathbb S^0$ and $\mathbb S^-$ could be shown to be universal. 

\section{Universality-centered Verifications}\label{further}

Let us thus set out to verify~(\ref{3-way}) and~(\ref{universal-studio}).  

First, given a partition ${\cal P}\in \mathscr P(N)$, define the patched-up cores as
\begin{equation}\label{patch-up}\begin{array}{l}
\mathbb X^{\mbox{i}\ast}(N,\vec v,{\cal P})\equiv \prod_{C\in {\cal P}}\mathbb X^\ast(C,\vec v|_{\mathscr C(C)})\\
\;\;\;\;\;\;\;\;\;\;\;\;\;\;\;\;\;\;\;\;=\left\{\vec x\equiv (x(i))_{i\in N}\in \Re^{N}:\;\vec x|_{C}\in \mathbb X^\ast(C,\vec v|_{\mathscr C(C)}),\;\;\;\forall C\in {\cal P}\right\},
\end{array}\end{equation}
where $\ast$ could be $+$, $0$, or $-$. By~(\ref{allocation-def}),~(\ref{core-def-trad}),~(\ref{core-def-weak}), and~(\ref{core-def-medium}) or~(\ref{core-def-middle}), $\mathbb X^{\mbox{i}\ast}(N,\vec v,{\cal P})$ of~(\ref{patch-up}) for any $\ast$ would be a subset of $\mathscr X(N,\vec v,{\cal P})$. Due to the peculiar nature of the all-consolidated partition $\{N\}$, as well as~(\ref{core-def-trad}),~(\ref{core-def-weak}),~(\ref{core-def-medium}), and~(\ref{patch-up}), $\mathbb X^{\mbox{i}\ast}(N,\vec v,\{N\})=\mathbb X^\ast(N,\vec v)$ for $\ast$ being $+$, $0$, or $-$. In other words, patched-up cores are extensions of ``plain'' cores from $\{N\}$-centric cases to more general cases catering to arbitrary partitions ${\cal P}$. 

Then, we define
\begin{equation}\label{fi-dominance}
\mathbb P^{\mbox{i}\ast}(N,\vec v)\equiv \left\{{\cal P}\in\mathscr P(N):\;\mathbb X^{\mbox{i}\ast}(N,\vec v,{\cal P})\neq\emptyset\right\},
\end{equation}
where $\ast$ could be $+$, $0$, or $-$. Built on~(\ref{patch-up}) and~(\ref{fi-dominance}), we shall show that
\begin{equation}\label{1-wayb}
\mathbb Q^{\mbox{i}\ast}(N,\vec v)=\bigcup_{{\cal P}\in \mathbb P^{\mbox{i}\ast}(N,\vec v)}\{{\cal P}\}\times \mathbb X^{\mbox{i}\ast}(N,\vec v,{\cal P}),
\end{equation} 
where $\ast$ could be $+$, $0$, or $-$. With~(\ref{weaker-core}),~(\ref{patch-up}), and~(\ref{fi-dominance}), the prospective~(\ref{1-wayb}) would lead to
\begin{equation}\label{georgia}
\mathbb Q^{\mbox{i}+}(N,\vec v)\subseteq \mathbb Q^{\mbox{i}0}(N,\vec v)\subseteq \mathbb Q^{\mbox{i}-}(N,\vec v).
\end{equation}
A combination of~(\ref{stab-def}) and~(\ref{georgia}) would then imply~(\ref{3-way}). 

Also, an alternative definition for $\mathbb P^{\mbox{i}0}(N,\vec v)$ of~(\ref{fi-dominance}) would be found to be
\begin{equation}\label{fi-dominance-alt}
\left\{{\cal P}\in\mathscr P(N):\;\tilde w(N,\vec v,{\cal P})\geq \tilde w(N,\vec v,{\cal P}'),\;\;\;\forall {\cal P}'\in \mathscr I(N,{\cal P})\right\}.
\end{equation}
For another definition 
\begin{equation}\label{fu-dominance}
\mathbb P^{\mbox{u}}(N,\vec v)\equiv \left\{{\cal P}\in\mathscr P(N):\;\tilde w(N,\vec v,{\cal P})\geq \tilde w(N,\vec v,{\cal P}'),\;\;\;\forall {\cal P}'\in \mathscr U(N,{\cal P})\right\},
\end{equation}
we shall show that
\begin{equation}\label{2-way}
\mathbb Q^{\mbox u}(N,\vec v)=\bigcup_{{\cal P}\in \mathbb P^{\mbox u}(N,\vec v)}\{{\cal P}\}\times \mathscr X(N,\vec v,{\cal P}).
\end{equation}

A combination of~(\ref{stab-def}),~(\ref{1-wayb}), and~(\ref{2-way}) would entail
\begin{equation}\label{stab-def-alternative}
\mathbb S^*(N,\vec v)=\bigcup_{{\cal P}\in \mathbb P^{\mbox{i}\ast}(N,\vec v)\cap \mathbb P^{\mbox{u}}(N,\vec v)}\mathbb X^{\mbox{i}\ast}(N,\vec v,{\cal P}),
\end{equation}
where $\ast$ could be $+$, $0$, or $-$; all the $\mathbb X^{\mbox{i}\ast}(N,\vec v,{\cal P})$'s for the ${\cal P}$'s that are allowed in the union sign are nonempty. But by~(\ref{fi-dominance-alt}) and~(\ref{fu-dominance}),
\begin{equation}\label{munia}
\mathbb P^{\mbox{i}0}(N,\vec v)\cap \mathbb P^{\mbox{u}}(N,\vec v)\neq\emptyset,
\end{equation}
as any partition ${\cal P}\in \mathscr P(N)$ that maximizes $\tilde w(N,\vec v,\cdot)$ must be a member of~(\ref{munia})'s left-hand side. Now~(\ref{stab-def-alternative}) and~(\ref{munia}) would together lead to~(\ref{universal-studio}). 

Thus, the verifications of~(\ref{3-way}) and~(\ref{universal-studio}) hinge on the proofs of~(\ref{1-wayb}),~(\ref{fi-dominance-alt}), and~(\ref{2-way}). 

To demonstrate~({\bf\ref{1-wayb}}), all we need is for $\ast$ being $+$, $0$, or $-$,
\begin{equation}\label{prop-2}
({\cal P},\vec x)\in \mathbb  Q^{\mbox{i}\ast}(N,\vec v)\;\;\;\Longleftrightarrow\;\;\;  (\{C\},\vec x|_{C})\in \mathbb Q^{\mbox{i}\ast}(C,\vec v|_{\mathscr C(C)}),\;\;\;\forall C\in {\cal P},
\end{equation}
thus linking a partition-allocation pair $({\cal P},\vec x)$'s various fission-resistant statuses to those of the pairs $(\{C\},\vec x|_C)$ made out of ${\cal P}$'s constituent coalitions $C$. To see this, note the second equality of~(\ref{badaling}) can be re-written as
\begin{equation}\label{pre-1wayb}
(\{N\},x)\in \mathbb Q^{\mbox{i}\ast}(N,\vec v)\;\;\;\Longleftrightarrow\;\;\; \vec x\in \mathbb X^*(N,\vec v).
\end{equation}
Combining~(\ref{patch-up}) with~(\ref{prop-2}) and~(\ref{pre-1wayb}), we may obtain
\begin{equation}\label{synonymous}
({\cal P},\vec x)\in \mathbb  Q^{\mbox{i}\ast}(N,\vec v)\;\;\;\Longleftrightarrow\;\;\; \vec x\in\mathbb X^{\mbox{i}\ast}(N,\vec v,{\cal P}).
\end{equation}
But in view of~(\ref{fi-dominance}), the above~(\ref{synonymous}) is just another way to express~(\ref{1-wayb}).  We now present the three cases for~(\ref{prop-2}). All proofs have been relegated to Appendix~\ref{app-a}. 

\begin{proposition}\label{prop-2-strong}
For any $\vec v\in \Re^{\mathscr C(N)}$, a partition-allocation pair $({\cal P}\equiv \{C_1,...,C_p\},\vec x)\in \mathscr Q(N,\vec v)$ of~(\ref{pair-def}) would be $(N,\vec v)$-{\bf strongly}-fission-resistant by Definition~\ref{def-fi-strong} if and only if each $(\{C_k\},\vec x|_{C_k})$ is $(C_k,\vec v|_{\mathscr C(C_k)})$-{\em strongly}-fission-resistant by Definition~\ref{def-fi-strong}. 
\end{proposition}

\begin{proposition}\label{prop-2-medium}
For any $\vec v\in \Re^{\mathscr C(N)}$, a partition-allocation pair $({\cal P}\equiv \{C_1,...,C_p\},\vec x)\in \mathscr Q(N,\vec v)$ of~(\ref{pair-def}) would be $(N,\vec v)$-{\bf mediumly}-fission-resistant by Definition~\ref{def-fi-medium} if and only if each $(\{C_k\},\vec x|_{C_k})$ is $(C_k,\vec v|_{\mathscr C(C_k)})$-{\em mediumly}-fission-resistant by Definition~\ref{def-fi-medium}. 
\end{proposition}

\begin{proposition}\label{prop-2-weak}
For any $\vec v\in \Re^{\mathscr C(N)}$, a partition-allocation pair $({\cal P}\equiv \{C_1,...,C_p\},\vec x)\in \mathscr Q(N,\vec v)$ of~(\ref{pair-def}) would be $(N,\vec v)$-{\bf weakly}-fission-resistant by Definition~\ref{def-fi-weak} if and only if each $(\{C_k\},\vec x|_{C_k})$ is $(C_k,\vec v|_{\mathscr C(C_k)})$-{\em weakly}-fission-resistant by Definition~\ref{def-fi-weak}. 
\end{proposition}

Proposition~\ref{prop-2-medium}, the medium version of~(\ref{prop-2}), could already be used to demonstrate~({\bf\ref{fi-dominance-alt}}). The proposition is about
\begin{equation}\label{prop-1-alt}\begin{array}{l}
\tilde w(N,\vec v,{\cal P})\geq \max_{{\cal P}'\in\mathscr I(N,\vec v)}\tilde w(N,\vec v,{\cal P}')\\
\hspace*{1.5in}\Longleftrightarrow\;\;\;v(C)\geq \max_{{\cal P}'\in \mathscr P(C)\setminus\{C\}}\tilde w(C,\vec v|_{\mathscr C(C)},{\cal P}'),\;\;\;\forall C\in {\cal P},
\end{array}\end{equation}
thus linking a ${\cal P}$'s dominance over its fission-down-to neighbors to those of all-consolidated partitions of its constituent coalitions. With~(\ref{core-def-medium}) and~(\ref{patch-up}), our~(\ref{prop-1-alt}) would entail
\begin{equation}\label{ganga}
\tilde w(N,\vec v,{\cal P})\geq \max_{{\cal P}'\in\mathscr I(N,\vec v)}\tilde w(N,\vec v,{\cal P}')\;\;\;\Longleftrightarrow\;\;\;\mathbb X^{\mbox{i}0}(N,\vec v,{\cal P})\neq\emptyset.
\end{equation}
In view of~(\ref{fi-dominance}), we see that~(\ref{ganga}) means~(\ref{fi-dominance-alt}). Indeed,~(\ref{allocation-def}) and~(\ref{core-def-medium}) could further lead to
\begin{equation}\label{meiyong}
\mathbb X^{\mbox{i}0}(N,\vec v,{\cal P})\neq\emptyset\;\;\;\Longleftrightarrow\;\;\;\mathbb X^{\mbox{i}0}(N,\vec v,{\cal P})=\mathscr X(N,\vec v,{\cal P}).
\end{equation}

%\begin{proposition}\label{last-minute}
%For any $\vec v\in \Re^{\mathscr C(N)}$, a partition ${\cal P}\equiv \{C_1....,C_p\}\in\mathscr P(N)$ would dominate its fission-down-to neighbors in $\mathscr I(N,\vec v)$ defined at~(\ref{fi-def}) in terms of the worth $\tilde w(N,\vec v,\cdot)$ of~(\ref{worth-def}) if and only if so does $\{C_k\}\in\mathbb P^{\mbox{i}}(C_k,\vec v|_{\mathscr C(C_k)})$ for every $k=1,...,p$.
%\end{proposition}

Meanwhile,~({\bf\ref{2-way}}) just amounts to
\begin{equation}\label{prop-4}
({\cal P},\vec x)\in \mathbb Q^{\mbox{u}}(N,\vec v)\Longleftrightarrow {\cal P}\in\mathbb P^{\mbox{u}}(N,\vec v),
\end{equation}
with the right-hand side placing no restriction on the allocation $\vec x\in\mathscr X(N,\vec v,{\cal P})$, thus linking a partition-allocation pair $({\cal P},\vec x)$'s fusion-resistant status to merely the dominance of ${\cal P}$ in worth over its fusion-up-to neighbors. 

\begin{proposition}\label{fu-equivalence}
For any $\vec v\in \Re^{\mathscr C(N)}$, a pair $({\cal P},\vec x)$ in $\mathscr Q(N,\vec v)$ of~(\ref{pair-def}) would be $(N,\vec v)$-fusion-resistant by Definition~\ref{def-fu} if and only if ${\cal P}\in \mathbb P^{\mbox{u}}(N,\vec v)$ of~(\ref{fu-dominance}).% while there is no extra requirement placed on $\vec x\in \mathscr X(N,\vec v,{\cal P})$ of~(\ref{allocation-def}).  
\end{proposition}

With~(\ref{1-wayb}),~(\ref{fi-dominance-alt}), and~(\ref{2-way}) thus confirmed, we have verified~(\ref{3-way}) and~(\ref{universal-studio}). Let us summarize our findings~(\ref{badaling}) to~(\ref{universal-studio}) in the following. 

\begin{theorem}\label{t-monumental}
The strong stability concept $\mathbb S^+$ based on Definitions~\ref{def-fi-strong} and~\ref{def-fu} is tighter than the medium stability concept $\mathbb S^0$ based on Definitions~\ref{def-fi-medium} and~\ref{def-fu}, which is in turn tighter than the weak stability concept $\mathbb S^-$ based on Definitions~\ref{def-fi-weak} and~\ref{def-fu}; each concept is compatible with its corresponding core concept defined at~(\ref{core-def-trad}),~(\ref{core-def-medium}), and~(\ref{core-def-weak}), respectively; importantly, the latter two stability concepts are universal.
\end{theorem}

\begin{figure}[h]
\centering
\includegraphics[width=1.1\columnwidth]{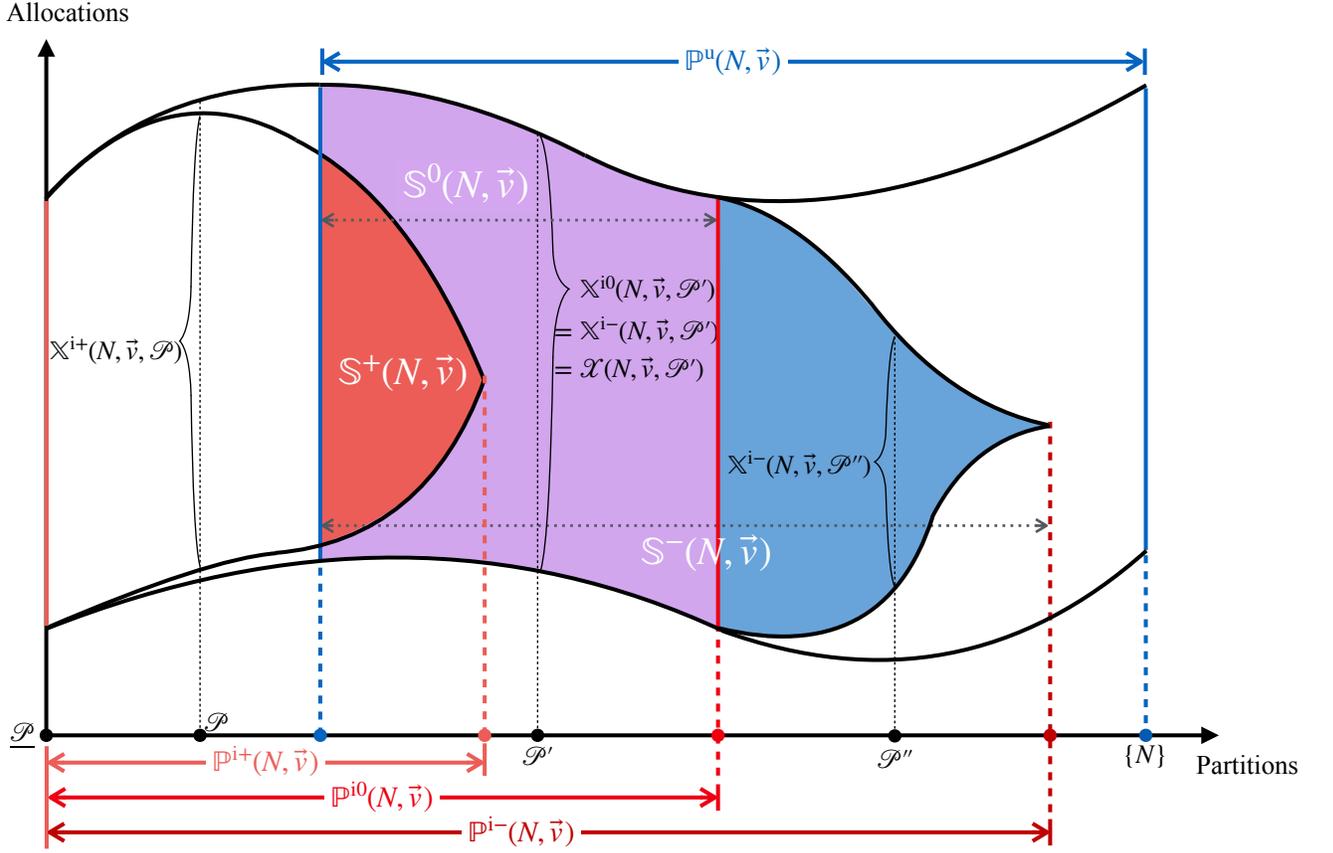}
\caption{A Schematic Sketch of the Various Entities}\label{rough-depiction}
\end{figure}

It is clear from~(\ref{fi-def}) and~(\ref{fu-def}) that $\mathscr I(N,{\cal P})\subset \mathscr I(N,{\cal P}')$ and $\mathscr U(N,{\cal P})\supset \mathscr U(N,{\cal P}')$ whenever ${\cal P}\prec {\cal P}'$. So by Definitions~\ref{def-fi-strong} to~\ref{def-fu}, it would be ``easier'' for a ``more splintered'' partition to be fission-resistant and a ``more consolidated'' partition to be fusion-resistant. Mindful of these, we use Figure~\ref{rough-depiction} to schematically sketch out the various entities. 

We are aware that $\mathscr X(N,\vec v,\underline{\cal P})$ for the all-splintered partition $\underline{\cal P}$ has exactly one member $\underline{\vec x}$ whose every component $\underline x(i)=v(\{i\})$; we are also aware that $\mathscr X(N,\vec v,{\cal P})$ could be empty for some partitions ${\cal P}$. Nevertheless, we have allowed all $\mathscr X(N,\vec v,{\cal P})$'s to have similar ``widths'' in Figure~\ref{rough-depiction} just for illustrative purposes. Moreover, the partitions are by no means totally ordered, nor are the sets $\mathbb P^{\mbox{i}+}(N,\vec v)$, $\mathbb P^{\mbox{i}0}(N,\vec v)$, $\mathbb P^{\mbox{i}-}(N,\vec v)$, and $\mathbb P^{\mbox{u}}(N,\vec v)$ necessarily connected. 

We have also drawn for a $\vec v$ with a nonempty $\mathbb S^+(N,\vec v)$. The latter could be empty for some other $\vec v$'s. On the other hand, Theorem~\ref{t-monumental} has ruled out the possibility of $\mathbb S^0(N,\vec v)$ ever being empty let alone $\mathbb S^-(N,\vec v)$. Under the current $\vec v$, even $\mathbb Q^{\mbox{i}-}(N,\vec v)$ has no intersection with $\{\{N\}\}\times \partial(N,\vec v)$. Thus, according to~(\ref{badaling}), even the weak core $\mathbb X^-(N,\vec v)$ is empty. On the flip side, some other $\vec v$'s could yet make the strong cores $\mathbb X^+(N,\vec v)$ nonempty. 

Suppose some $\mathbb Q^{\mbox{i}\ast}(N,\vec v)$ is so ``extensive'' as to contain the ``rightmost'' portion of Figure~\ref{rough-depiction} corresponding to the all-consolidated partition $\{N\}$. Then, as $\{\{N\}\}\times \partial (N,\vec v)$ is surely a part of $\mathbb Q^{\mbox u}(N,\vec v)$, this would be the case where a nonempty $\{\{N\}\}\times \mathbb X^*(N,\vec v)$ is contained in $\mathbb S^*(N,\vec v)$. Here, indeed, all three sets involved in~(\ref{badaling}) would be simultaneously nonempty. 

By~(\ref{fi-dominance}),~(\ref{fi-dominance-alt}), and~(\ref{fu-dominance}), $({\cal P},\vec x)\in \mathbb S^0(N,\vec v)$ of~(\ref{stab-def-alternative}) if and only if\\
\indent\M (hp) ${\cal P}\in \mathbb P^{\mbox{i}0}(N,\vec v)\cap \mathbb P^{\mbox{u}}(N,\vec v)$ or in the words of Apt and Radzik \cite{AR06}, is $\mathbb D_{hp}$-stable; \\
\indent\M (x) also, $\vec x\in \mathbb X^{\mbox{i}0}(N,\vec v,{\cal P})$ of~(\ref{patch-up}).

The stability notions of Apt and Radzik \cite{AR06} did not specify the (x) portion; namely, how individual payers are allotted their shares. According to Example~\ref{example-a}, however, such allocations matter even for the special case where ${\cal P}=\{N\}$: when the traditional core $\mathbb X^+(N,\vec v)$ is mis-used there for $\mathbb X^0(N,\vec v)\equiv \mathbb X^{\mbox{i}0}(N,\vec v,{\cal P}=\{N\})$, no stable solution need exist.  

\section{Road to Medium Stability}\label{pathway}

There is a natural process to move from any given partition ${\cal P}\in \mathscr P(N)$ on to a terminal partition ${\cal P}^0\in \mathbb P^{\mbox{i}0}(N,\vec v)\cap \mathbb P^{\mbox{u}}(N,\vec v)$ and then a pair $({\cal P}^0,\vec x^{\,0})$ in the mediumly stable set $\mathbb S^0(N,\vec v)$. This process can best be described as a number of moves in a graph.  

Note $\mathscr P(N)$ is decomposable into $n$ subsets $\mathscr P^{1}(N),\mathscr P^2(N),...,\mathscr P^n(N)$ with each $\mathscr P^p(N)$ containing partitions ${\cal P}\in\mathscr P(N)$ such that $|{\cal P}|=p$. Clearly, $\mathscr P^1(N)$ contains but one member, the all-consolidated $\{N\}$; so does $\mathscr P^n(N)$ with its sole member being the all-splintered $\underline{\cal P}$. For a $p$ in between, its $\mathscr P^p(N)$ could be bigger. 

\begin{figure}[h]
\centering
\includegraphics[width=1.0\columnwidth]{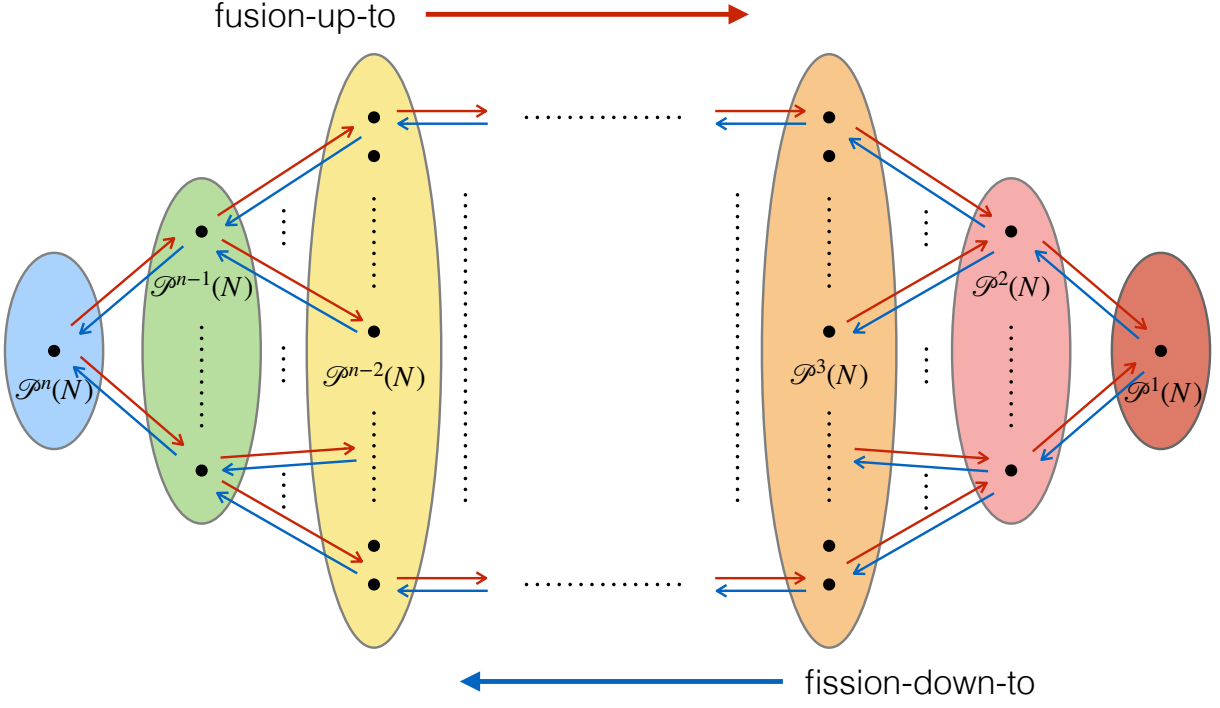}
\caption{A Depiction of the $n$-partite Graph $\mathscr G(N)$}\label{depiction}
\end{figure}

Treating the $\mathscr P^p(N)$'s as sets of nodes, we can obtain an $n$-partite graph
\[ \mathscr G(N)\equiv (\mathscr P^n(N),\mathscr A^{n-1,n}(N),\mathscr A^{n,n-1}(N),\mathscr P^{n-1}(N),...,\mathscr P^{2}(N),\mathscr A^{1,2}(N),\mathscr A^{2,1}(N),\mathscr P^1(N)),\]
where each $\mathscr A^{p,p+1}(N)$ contains all ``one-step fission'' arcs from $\mathscr P^p(N)$ to $\mathscr P^{p+1}(N)$ and each $\mathscr A^{p+1,p}(N)$ contains all ``one-step fusion'' arcs from $\mathscr P^{p+1}(N)$ to $\mathscr P^{p}(N)$. Each one-step fission/fusion involves merely one starting/ending coalition and a two-way split/merger from/into it.  
%Arcs $({\cal P},{\cal P}')$ in both $\mathscr A^{p,p+1}(N)$ and $\mathscr A^{p+1,p}(N)$ are obviously defined. 
%Any one-step fission arc ${\cal A}\equiv ({\cal P},{\cal P}')\in  \mathscr A^{p,p+1}(N)$ would have ${\cal P}\in \mathscr P^p(N)$ and ${\cal P}'\in \mathscr P^{p+1}(N)$; also, ${\cal P}$ is representable as $\{C_1,...,C_{p-1},C_p\}$ and ${\cal P}'$ as $\{C_1,...,C_{p-1},C'_{p},C'_{p+1}\}$ such that $C'_p$ and $C'_{p+1}$ form a partition of $C_p$. Any one-step fusion arc ${\cal A}\equiv ({\cal P},{\cal P}')\in  \mathscr A^{p+1,p}(N)$ would have ${\cal P}\in \mathscr P^{p+1}(N)$ and ${\cal P}'\in \mathscr P^{p}(N)$; also, ${\cal P}$ is representable as $\{C_1,...,C_{p-1},C_p,C_{p+1}\}$ and ${\cal P}'$ as $\{C_1,...,C_{p-1},C_{p}\cup C_{p+1}\}$. 
%Clearly, $({\cal P},{\cal P}')\in \mathscr A^{p+1,p}(N)$ if and only if $({\cal P}',{\cal P})\in \mathscr A^{p,p+1}(N)$. 
Figure~\ref{depiction} is a depiction of the graph $\mathscr G(N)$. 

It is clear from Figure~\ref{depiction} that $\mathscr G(N)$ is the directed-graph version of what Sandholm et al. \cite{SLAST99} would call a coalition structure graph. Our directed arcs would help distinguish between the fission-down-to and fusion-up-to directions. More details about the graph $\mathscr G(N)$, including its connectivity, have been provided in Appendix~\ref{app-ab}.  

The following steepest ascent method (SAM), while implementable by moves in the graph, can ensure the finite-step convergence to a point in $\mathbb S^0(N,\vec v)$:\\
\vspace*{.04in}\indent\M (m) Keep on moving from any current ${\cal P}$ to a new ${\cal P}'$ as long as $\tilde w(N,\vec v,{\cal P})<\tilde w(N,\vec v,{\cal P}')$ with ${\cal P}'\in\mbox{argmax}_{\;{\cal P}''\in \mathscr I(N,{\cal P})\cup\mathscr U(N,{\cal P})}\tilde w(N,\vec v,{\cal P}'')$ 
until such a move is no longer possible.\\
\vspace*{.04in}\indent\M (s) The current ${\cal P}^0$ must belong to both $\mathbb P^{\mbox{i}0}(N,\vec v)$ of~(\ref{fi-dominance}) or~(\ref{fi-dominance-alt}) and $\mathbb P^{\mbox{u}}(N,\vec v)$ of~(\ref{fu-dominance}). Then, pick any $\vec x^{\,0}|_{C}\in \mathbb X^0(C,\vec v|_{\mathscr C(C)})\equiv \partial(C,\vec v|_{\mathscr C(C)})$ for each $C\in {\cal P}^0$ and assemble them into a whole $\vec x^{\,0}\in\mathbb X^{\mbox{i}0}(N,\vec v,{\cal P}^0)\equiv \mathscr X(N,\vec v,{\cal P}^0)$.  

\vspace*{.04in}As indicated by Appendix~\ref{app-ab}, each of SAM's (m)-steps consists of a number of moves in the graph $\mathscr G(N)$, which is finite and also connected. Since each (m)-step ensures that $\tilde w(N,\vec v,\cdot)$ strictly increases, (SAM) would reach its (s)-step from any starting ${\cal P}$ in a finite number of (m)-steps. The membership of the terminal ${\cal P}^0$ to $\mathbb P^{\mbox{i}0}(N,\vec v)\cap \mathbb P^{\mbox{u}}(N,\vec v)$ is clear from~(\ref{fi-dominance-alt}) and~(\ref{fu-dominance}). Our search for $\vec x^{\,0}$ to make up for a partition-allocation pair $({\cal P}^0,\vec x^{\,0})\in\mathbb S^0(N,\vec v)$ is guaranteed to succeed by~(\ref{core-def-medium}),~(\ref{patch-up}), and~(\ref{fi-dominance}). 

SAM is by no means an always one-shot deal. For instance, it could go from $\{\{1,2\},\{3\}\}$ to $\{\{1\},\{2\},\{3\}\}$, and then on to $\{\{1\},\{2,3\}\}$. We speculate how well the evolution of partitions during one SAM run might mimic a real process, where players might be individual lions on the Savannah, lawyers in a city, or administrative entities on a continent. %We leave potential simulations of such ethological, economic, or political situations to future research. 
Some other details are again left to Appendix~\ref{app-ab}.

At a particular partition ${\cal P}\equiv \{C_1,...,C_p\}$ while running SAM during the (m) steps, deciding whether ${\cal P}\in \mathbb P^{\mbox{i}0}(N,\vec v)\cap\mathbb P^{\mbox{u}}(N,\vec v)$ would boil down to computing the various worths $\tilde w(N,\vec v,{\cal P}')$ defined through~(\ref{worth-def}) for ${\cal P}'\in \{{\cal P}\}\cup\mathscr I(N,{\cal P})\cup\mathscr U(N,{\cal P})$ and comparing them. Since the neighborhoods defined at~(\ref{fi-def}) and~(\ref{fu-def}) could be combinatorial in size, proper integer programming (IP) or dynamic programming (DP) formulation might help. One may consult relevant works such as Rahman et al. \cite{RMWJ15} and Sandholm et al. \cite{SLAST99}.  

The (s) step concerning a terminal partition ${\cal P}^0$ involves $|{\cal P}^0|$ incidences. For each $C\in {\cal P}^0$, it is about identifying one $\vec x^{\,0}|_C$ out of the medium core $\mathbb X^0(C,\vec v|_{\mathscr C(C)})\equiv \partial(C,\vec v|_{\mathscr C(C)})$ that is known to be nonempty. By~(\ref{eff-allocation}), linear programming (LP) would be a good tool for this.

Since $({\cal P}^0,\vec x^{\,0})\in \mathbb S^0(N,\vec v)\subseteq \mathbb S^-(N,\vec v)$, SAM's final output $({\cal P}^0,\vec x^{\,0})$ can already resist temptations unless they are too strong. First, there would be no incentive for any constituent coalitions $C$ in ${\cal P}^0$ to merge further. Second, for any such $C$, any attempt on its further split would have to overcome at least one roadblock planted by a disgruntled sub-coalition. 

\section{Concluding Remarks}\label{conclusion}

We have extended the traditional core concept to its looser brethrens, the medium and the weak cores. Even the last one can be rationalized via the principle of consensual blocking---to succeed in splitting the grand coalition it would take the consensus of all affected parties. In this light, the traditonal core can be cast as one that condones unilateral blocking. With the conviction that most if not all coalitional-game incidences are supposed to have their ``natural resting states'' where players by no means all end up in the grand coalition, we have applied these principles to partition-allocation pairs.   

As a result, we have achieved three stability concepts, the strong, the medium, and the weak. Strong stability is already more inclusive than its traditional-core counterpart; meanwhile, medium and weak stabilities have clinched the trophy of universality: every game, no mater how ``poor'' it is, has its fair share of stable solutions. A greedy algorithm for mediumly stable solutions has been proposed. It will be interesting to see whether real evolutions could have semblances to the algorithm's running process. 

We also wonder whether similar extensions can be made to more general or refined notions such as games in the partition function form as proposed by Thrall and Lucas \cite{TL63} or the population monotonic allocation schemes as proposed by Sprumont \cite{S90}. 
%It is hoped that our SAM for reaching local stability would offer some semblance to real-world evolutions of alliances in search of more stability. But in the real world, coalition values change over time and when participants make alliance-forming decisions, future prospects must appear random to them. Besides, entry or exit charges might be factors to be reckoned with as well. In any event, there is still a long way to go before an academic model like ours could start to resemble reality to a reasonably degree. Furthermore, for the IP formulations proposed in the paper, we will have to leave their comparative empirical studies, complexity analyses, and potential improvements to the future.    
As the traditional core is tightly associated with balancedness, we naturally question whether the weak core is related to some kind of weak balancedness. %The fact that~(\ref{p-mip}) is an MILP rather than a pure LP might have complicated the matter. 
Moreover, our universal stability notions are compatible with the medium and weak cores but not the traditional core. It would be good to know whether there are core-compatible stability notions that still uphold universality. 

\vspace*{.5in}

\noindent{\bf\Large Acknowledgments}

\vspace*{.05in}The author would like to thank two referees of an earlier version of this paper. They raised counter examples, quite similar to Example~\ref{example-a} here, which refuted an erroneous claim made earlier. The current concepts related to the weak core are attributable to this impetus.  

\vspace*{.5in}

\noindent{\bf\Large Appendices}%\vspace*{.05in}
\appendix
\numberwithin{equation}{section}

\section{Technical Details of Section~\ref{further}}\label{app-a}

\noindent{\bf Proof of Proposition~\ref{prop-2-strong}: }For the ``if'' part, suppose each $(\{C_k\},\vec x|_{C_k})$ is $(C_k,\vec v|_{\mathscr C(C_k)})$-strongly-fission-resistant. Any ${\cal P}'\in\mathscr I(N,{\cal P})$ must take the form $\{C^1_1,...,C^{l_1}_1,...,C^1_p,...,C^{l_p}_p\}$ such that for each $k=1,...,p$, ${\cal P}'_k\equiv\{C^1_k,...,C^{l_k}_k\}$ is a partition of $C_k$ and hence either just $\{C_k\}$ itself or in the latter's fission-down-to neighborhood $\mathscr I(C_k,\{C_k\})$. 

For any $\vec x^{\;\prime}\in \mathscr X(N,\vec v,{\cal P}')$ and $k=1,...,p$, we have by~(\ref{allocation-def}) that $\vec x^{\;\prime}|_{C_k}\in \mathscr X(C_k,\vec v|_{\mathscr C(C_k)},{\cal P}'_k)$. When $l_k=1$, both $\vec x|_{C_k}$ and $\vec x^{\;\prime}|_{C_k}$ are members of $\mathscr X(C_k,\vec v|_{\mathscr C(C_k)},\{C_k\})$. By~(\ref{allocation-def}),
\[ \sum_{i\in C_k}x(i)=v(C_k)=\sum_{i\in C_k}x'(i).\]
When $l_k\geq 2$ and hence ${\cal P}'_k\equiv \{C^1_k,...,C^{l_k}_k\}\in\mathscr I(C_k,\{C_k\})$, on the other hand, we have from the hypothesis and Definition~\ref{def-fi-strong} that 
\[ \sum_{i\in C^j_k}x(i)\geq \sum_{i\in C^j_k}x'(i),\hspace*{.8in}\forall j=1,...,l_k.\]
By Definition~\ref{def-fi-strong}, $({\cal P},\vec x)$ must be $(N,\vec v)$-strongly-fission-resistant. 

For the ``only if'' part, suppose the pair formed by a partition ${\cal P}\equiv \{C_1,...,C_p\}$ and one of its allocation vectors $\vec x\in \mathscr X(N,\vec v,{\cal P})$ is $(N,\vec v)$-strongly-fission-resistant. For any $k=1,...,p$, let $\{C^1_k,...,C^l_k\}\in \mathscr I(C_k,\{C_k\})$. Now ${\cal P}'\equiv \{C_1,...,C_{k-1},C^1_k,...,C^l_k,C_{k+1},...,C_p\}$ is a member of $\mathscr I(N,{\cal P})$. For any $\vec x^{\;\prime}\in \mathscr X(C_k,\vec v|_{\mathscr C(C_k)},\{C^1_k,...,C^l_k\})$, note by~(\ref{allocation-def}) that the concatenation $\vec x^{\;\prime\prime}\in \Re^N$ with $\vec x^{\;\prime\prime}|_{C_j}=\vec x|_{C_j}$ for $j=1,...,k-1,k+1,...,p$ and $\vec x^{\;\prime\prime}|_{C^j_k}=\vec x^{\;\prime}|_{C^j_k}$ for $j=1,...,l$ would be in $\mathscr X(N,\vec v,{\cal P}')$. But by the hypothesis and Definition~\ref{def-fi-strong},
\[ \sum_{i\in C^j_k}x(i)\geq \sum_{i\in C^j_k}x''(i)=\sum_{i\in C^j_k}x'(i),\hspace*{.8in}\forall j=1,...,l.\]
The above would mean the $(C_k,\vec v|_{\mathscr C(C_k)})$-strongly-fission-resistance status of $(\{C_k\},\vec x|_{C_k})$. % by Definition~\ref{def-fi}. 
\qed

\noindent{\bf Proof of Proposition~\ref{prop-2-medium}: }For the ``if'' part, suppose every $\{C_k\}$ dominates all further partitions. Any ${\cal P}'\in\mathscr I(N,{\cal P})$ of~(\ref{fi-def}) must be of the form $\{C^1_1,...,C^{l_1}_1,...,C^1_p,...,C^{l_p}_p\}$ where for each $k=1,...,p$, we have ${\cal P}'_k\equiv \{C^1_k,...,C^{l_k}_k\}$ being a member of $\mathscr I(C_k,\{C_k\})$. Hence, 
\[\begin{array}{l}
\tilde w(N,\vec v,{\cal P})=v(C_1)+\cdots+v(C_p)=\tilde w(C_1,\vec v|_{\mathscr C(C_1)},\{C_1\})+\cdots+\tilde w(C_p,\vec v|_{\mathscr C(C_p)},\{C_p\})\\
\;\;\;\;\;\;\;\;\;\;\;\;\;\;\;\;\;\geq \tilde w(C_1,\vec v|_{\mathscr C(C_1)},{\cal P}'_1)+\cdots+\tilde w(C_p,\vec v|_{\mathscr C(C_p)},{\cal P}'_p)\\
\;\;\;\;\;\;\;\;\;\;\;\;\;\;\;\;\;=\sum_{j=1}^{l_1}v(C^j_1)+\cdots+\sum_{j=1}^{l_p}v(C^j_{p})=\tilde w(N,\vec v,{\cal P}'),
\end{array}\]
where the first equality is just~(\ref{worth-def}), the second equality stems from the same and the peculiar natures of the all-consolidated $\{C_k\}$'s, the inequality follows the current hypothesis involving Definition~\ref{def-fi-medium}, the third equality is due to~(\ref{worth-def}) and each ${\cal P}'_k$'s formation, and the last equality follows from~(\ref{worth-def}) and ${\cal P}'$'s formation. By~(\ref{fi-dominance-alt}), this means that ${\cal P}\in \mathbb P^{\mbox{i}0}(N,\vec v)$; equivalently, it means $({\cal P},\vec x)$'s satisfaction of Definition~\ref{def-fi-medium} for $(N,\vec v)$. 

For the ``only if'' part, suppose ${\cal P}$ dominates all its fission-down-to neighbors in worth. For any $k=1,...,p$, let ${\cal P}'_k\equiv \{C^1_k,...,C^l_k\}$ be a partition of $C_k$. Consider ${\cal P}''\equiv \{C_1,...,C_{k-1},C^1_k,...,$ $C^l_k,C_{k+1},...,C_p\}$ which is a member of $\mathscr I(N,\vec v,{\cal P})$. Then, 
\[\begin{array}{l}
\tilde w(C_k,\vec v|_{\mathscr C(C_k)},\{C_k\})=v(C_k)=\tilde w(N,\vec v,{\cal P})-v(C_1)-\cdots-v(C_{k-1})-v(C_{k+1})-\cdots-v(C_p)\\
\;\;\;\;\;\;\;\;\;\;\;\;\;\;\;\;\;\;\;\;\;\;\;\;\;\;\;\;\;\;\;\;\geq\tilde w(N,\vec v,{\cal P}'')-v(C_1)-\cdots-v(C_{k-1})-v(C_{k+1})-\cdots-v(C_p)\\
\;\;\;\;\;\;\;\;\;\;\;\;\;\;\;\;\;\;\;\;\;\;\;\;\;\;\;\;\;\;\;\;=v(C^1_k)+\cdots+v(C^l_k)=\tilde w(C_k,\vec v|_{\mathscr C(C_k)},{\cal P}'_k),
\end{array}\] 
where the first two equalities both stem from~(\ref{worth-def}), as well as the definitions of $\{C_k\}$ and ${\cal P}$, the inequality follows from the hypothesis, the third and fourth equalities are again due to~(\ref{worth-def}), as well as the definitions of ${\cal P}''$ and ${\cal P}'_k$. Hence, $\{C_k\}\in\mathbb P^{\mbox{i}0}(C_k,\vec v|_{\mathscr C(C_k)})$ by~(\ref{fi-dominance-alt}); in other words, $(\{C_k\},\vec x|_{C_k})$ satisfies Definition~\ref{def-fi-medium} for $(C_k,\vec v|_{\mathscr C(C_k)})$ . \qed

\noindent{\bf Proof of Proposition~\ref{prop-2-weak}: }For the ``if'' part, suppose each $(\{C_k\},\vec x|_{C_k})$ is $(C_k,\vec v|_{\mathscr C(C_k)})$-weakly-fission-resistant. Note a ${\cal P}'\in\mathscr I(N,{\cal P})$ must be of the form $\{C^1_1,...,C^{l_1}_1,...,C^1_p,...,C^{l_p}_p\}$ such that for each $k=1,...,p$, ${\cal P}'_k\equiv\{C^1_k,...,C^{l_k}_k\}$ is a partition of $C_k$ and hence either just $\{C_k\}$ itself or in the latter's fission-down-to neighborhood $\mathscr I(C_k,\{C_k\})$.  

For any $\vec x^{\;\prime}\in \mathscr X(N,\vec v,{\cal P}')$ and $k=1,...,p$, we have by~(\ref{allocation-def}) that $\vec x^{\;\prime}|_{C_k}\in \mathscr X(C_k,\vec v|_{\mathscr C(C_k)},{\cal P}'_k)$. As ${\cal P}'\in \mathscr I(N,{\cal P})$ of~(\ref{fi-def}) is not ${\cal P}$ itself, we must have $l_k\geq 2$ and hence ${\cal P}'_k\in \mathscr I(C_k,\{C_k\})$ for at least one $k=1,...,p$. For this $k$, we have from the hypothesis and Definition~\ref{def-fi-weak} that 
\[	\sum_{i\in C^j_k}x(i)\geq \sum_{i\in C^j_k}x'(i),\hspace*{.8in}\exists j=1,...,l_k.\]
%When $l_k=1$, both $\vec x|_{C_k}$ and $\vec x^{\;\prime}|_{C_k}$ are members of $\mathscr X(C_k,\vec v|_{\mathscr C(C_k)},\{C_k\})$. By~(\ref{allocation-def}),
%\[ \sum_{i\in C_k}x(i)=v(C_k)=\sum_{i\in C_k}x'(i).\]
%When $l_k\geq 2$ and hence ${\cal P}'_k\equiv \{C^1_k,...,C^{l_k}_k\}\in\mathscr I(C_k,\{C_k\})$, 
Thus, by Definition~\ref{def-fi-weak}, $({\cal P},\vec x)$ must be $(N,\vec v)$-weakly-fission-resistant. 

For the ``only if'' part, suppose the pair formed by a partition ${\cal P}\equiv \{C_1,...,C_p\}$ and one of its allocation vectors $\vec x\in \mathscr X(N,\vec v,{\cal P})$ is $(N,\vec v)$-weakly-fission-resistant. For any $k=1,...,p$, let $\{C^1_k,...,C^l_k\}\in \mathscr I(C_k,\{C_k\})$ and hence $l\geq 2$. Now ${\cal P}'\equiv \{C_1,...,C_{k-1},C^1_k,...,C^l_k,C_{k+1},...,C_p\}$ is a member of $\mathscr I(N,{\cal P})$. For any $\vec x^{\;\prime}\in \mathscr X(C_k,\vec v|_{\mathscr C(C_k)},\{C^1_k,...,C^l_k\})$, note by~(\ref{allocation-def}) that the concatenation $\vec x^{\;\prime\prime}\in \Re^N$ with $\vec x^{\;\prime\prime}|_{C_j}=\vec x|_{C_j}$ for $j=1,...,k-1,k+1,...,p$ and $\vec x^{\;\prime\prime}|_{C^j_k}=\vec x^{\;\prime}|_{C^j_k}$ for $j=1,...,l$ would be in $\mathscr X(N,\vec v,{\cal P}')$. But by the hypothesis, Definition~\ref{def-fi-weak}, and the fact that $C_1$, $...$, $C_{k-1}$, $C_{k+1}$, $...$, $C_p$ all belong to ${\cal P}$ and hence do not count as the $C'$ in Definition~\ref{def-fi-weak},
\[ \sum_{i\in C^j_k}x(i)\geq \sum_{i\in C^j_k}x''(i)=\sum_{i\in C^j_k}x'(i),\hspace*{.8in}\exists j=1,...,l.\]
The above would mean the $(C_k,\vec v|_{\mathscr C(C_k)})$-weak-fission-resistance status of $(\{C_k\},\vec x|_{C_k})$. % by Definition~\ref{def-fi}. 
\qed

\noindent{\bf Proof of Proposition~\ref{fu-equivalence}: }For the ``if'' part, suppose ${\cal P}\equiv \{C_1,...,C_p\}$ dominates all fusion-up-to neighbors in terms of $\tilde w(N,\vec v,\cdot)$ defined at~(\ref{worth-def}). Let ${\cal P}'\equiv \{C'_1,...,C'_{p'}\}$ be such a neighbor. The coalition $C'_k$ for any $k=1,...,p'$ must be a union of some coalitions $C_l$ that make up ${\cal P}$. Let us construct ${\cal P}''\in \mathscr U(N,{\cal P})$ of~(\ref{fu-def}) by putting this particular $C'_k$ together with all ${\cal P}$-components $C_{l}$ that do not intersect with $C'_k$. Now
\[\begin{array}{l}
\sum_{l:\;C_l\cap C'_k\neq\emptyset}v(C_l)=\sum_{l=1}^pv(C_l)-\sum_{l:\;C_l\cap C'_k=\emptyset}v(C_l)=\tilde w(N,\vec v,{\cal P})-\sum_{l:\;C_l\cap C'_k=\emptyset}v(C_l)\\
\;\;\;\;\;\;\;\;\;\;\;\;\;\;\;\;\;\;\;\;\;\;\;\;\;\;\;\geq \tilde w(N,\vec v,{\cal P}'')-\sum_{l:\;C_l\cap C'_k=\emptyset}v(C_l)=v(C'_k),
\end{array}\] 
where the first equality is just an identity, the second equality stems from~(\ref{worth-def}), the inequality is due to the ${\cal P}$-dominance hypothesis in the sense of~(\ref{fu-dominance}), the last equality comes from~(\ref{worth-def}) and the way ${\cal P}''$ is constructed. By Definition~\ref{def-fu}, we know $({\cal P},\vec x)$ is $(N,\vec v)$-fusion-resistant. 

For the ``only if'' part, suppose $({\cal P}\equiv \{C_1,...,C_p\},\vec x)$ is $(N,\vec v)$-fusion-resistant. Let ${\cal P}'\equiv \{C'_1,...,C'_{p'}\}$ be an arbitrary member of $\mathscr U(N,{\cal P})$. Note each $C'_k$ for $k=1,...,p'$ would be exactly the union of those $C_l$'s that intersect with it. Now
\[ \tilde w(N,\vec v,{\cal P})=\sum_{l=1}^pv(C_l)=\sum_{k=1}^{p'}\;\sum_{l:\;C_l\cap C'_k\neq\emptyset}v(C_l)\geq \sum_{k=1}^{p'}v(C'_k)=\tilde w(N,\vec v,{\cal P}'),\]
where the first equality follows from~(\ref{worth-def}), the second equality is just an identity, the inequality comes from the hypothesis, Definition~\ref{def-fu}, and the preservation of inequalities under summation, and the last equality is again due to~(\ref{worth-def}). Therefore, ${\cal P}\in\mathbb P^{\mbox{u}}(N,\vec v)$ of~(\ref{fu-dominance}).\qed

\section{Technical Details of Section~\ref{pathway}}\label{app-ab}

\noindent{\bf Details linked to $\mathscr G(N)$: }Let us denote $({\cal P},{\cal P}')\in \mathscr A^{p+1,p}(N)$ or the equivalent $({\cal P}',{\cal P})\in \mathscr A^{p,p+1}(N)$ by ${\cal P}\prec_1{\cal P}'$. For ${\cal P}\equiv \{C_1,...,C_p\}\in \mathscr P^p(N)$ with $p=1,2,...,n-1$, we call
\begin{equation}\label{fission-to}
\mathscr I_1(N,{\cal P})\equiv \{{\cal P}'\in \mathscr P^{p+1}(N):\;({\cal P},{\cal P}')\in \mathscr A^{p,p+1}(N)\mbox{ or equivalently }{\cal P}'\prec_1{\cal P}\}
\end{equation}
its one-step fission-down-to neighborhood. We can always let $\mathscr I_1(N,\{\{i_1\},...,\{i_n\}\})=\emptyset$. For ${\cal P}\equiv \{C_1,...,C_p\}\in \mathscr P^p(N)$ with $p=n,n-1,...,2$, we call
\begin{equation}\label{fusion-to}
\mathscr U_1(N,{\cal P})\equiv \{{\cal P}'\in \mathscr P^{p-1}(N):\;({\cal P},{\cal P}')\in \mathscr A^{p,p-1}(N)\mbox{ or equivalently }{\cal P}\prec_1{\cal P}'\}
\end{equation}
its one-step fusion-up-to neighborhood. We can always let $\mathscr U_1(N,\{N\})=\emptyset$. %It is easy to reach $|\mathscr U^p_1(N,{\cal P})|=p\cdot (p-1)/2$ for any $p$.

As any fission/fusion is implementable as a number of one-step  fission/fusion moves, the following should be immediate from~(\ref{fi-def}),~(\ref{fu-def}),~(\ref{fission-to}), and~(\ref{fusion-to}). For $p,p'=1,2,...,n$ with $p\geq p'+1$, suppose ${\cal P}\in \mathscr P^p(N)$ and ${\cal P}'\in\mathscr P^{p'}(N)$. Then,   
\begin{equation}\label{thoroughgoing}
{\cal P}\in \mathscr I(N,{\cal P}')\Longleftrightarrow {\cal P}\prec{\cal P}'\Longleftrightarrow \mathscr U(N,{\cal P})\ni {\cal P}'
\end{equation}
would be true if and only if there are ${\cal P}_p\equiv {\cal P}$, ${\cal P}_{p-1}$, $\cdots$, ${\cal P}_{p'+1}$, ${\cal P}_{p'}\equiv {\cal P}'$ such that for any $k=p-1,p-2,...,p'$, 
\begin{equation}\label{one-step}
{\cal P}_{k+1}\in \mathscr I_1(N,{\cal P}_k)\Longleftrightarrow {\cal P}_{k+1}\prec_1{\cal P}_k\Longleftrightarrow \mathscr U_1(N,{\cal P}_{k+1})\ni {\cal P}_k.
\end{equation}
In view of~(\ref{thoroughgoing}) and~(\ref{one-step}), $\mathscr I(N,{\cal P})$ may be understood as the entire branch to the left of ${\cal P}$ reachable via one-step fission-down-to arcs; meanwhile, $\mathscr U(N,{\cal P})$ may be understood as the entire branch to the right of ${\cal P}$ reachable via one-step fusion-up-to arcs.

We are sure that any part of the graph is reachable from its any other part.  

\begin{fact}\label{connectivity}
The $n$-partite graph $\mathscr G(N)$ is connected.
\end{fact}

\vspace*{-.1in}\noindent{\bf Proof: }For any ${\cal P}\equiv \{C_1,...,C_p\}\in\mathscr P^p(N)\subseteq \mathscr P(N)$ and ${\cal P}'\equiv \{C'_1,...,C'_{p'}\}\in\mathscr P^{p'}(N)\subseteq \mathscr P(N)$, consider
\[ {\cal P}''\equiv {\cal P}\wedge_{(\prec)} {\cal P}'\equiv \{C''_1,...,C''_{p''}\}\in \mathscr P^{p''}(N)\subseteq \mathscr P(N),\]
such that $p''\geq p\vee p'$ and each $C''_{k''}$ is the nonempty intersection of some $C_{k}$ and $C'_{k'}$. Note
\[ {\cal P}''\in \mathscr I(N,{\cal P})\;\;\;\;\;\;\mbox{ and yet }\;\;\;\;\;\;{\cal P}'\in \mathscr U(N,{\cal P}'').\]
Due to the link between~(\ref{thoroughgoing}) and~(\ref{one-step}), we can go from ${\cal P}$ to ${\cal P}'$  via ${\cal P}''$ in a finite number of steps. Symmetrically, our proof can be fulfilled by letting ${\cal P}$ go up to ${\cal P}''\equiv {\cal P}\vee_{(\prec)}{\cal P}'$ through fusions first and then going down to ${\cal P}'$ through fissions. \qed 

\noindent{\bf Other Details concerning SAM: }The worth $\tilde w(N,\vec v,{\cal P})$ is clearly on the rise during a SAM run. The increases actually occur at even finer levels. %need not specify the allocations $\vec x$ during intermediate steps---its step (s) is needed only in the very end. In a real process, however, there must be migrations from pairs $({\cal P},\vec x)$ to pairs $({\cal P}',\vec x^{\;\prime})$, where the allocations belong to the partition-corresponding patched-up-cores. We can show that players would be well-motivated in every step of such a process.

\begin{fact}\label{p-motivation}
Suppose partitions ${\cal P}\equiv \{C_1,...,C_p\}$ and ${\cal P}'\equiv \{C^1_1,...,C^{l_1}_1,...,C^1_p,...,C^{l_p}_p\}$ satisfy ${\cal P}'\in\mbox{argmax}_{\;{\cal P}''\in \mathscr I(N,{\cal P})}\tilde w(N,\vec v,{\cal P}'')$, then $\tilde w(N,\vec v,{\cal P}')\geq \tilde w(N,\vec v,{\cal P})$ if and only if 
\[ \tilde w\left(C_k,\vec v|_{\mathscr C(C_k)},\{C^1_k,...,C^{l_k}_k\}\right)\geq \tilde w\left(C_k,\vec v|_{\mathscr C(C_k)},\{C_k\}\right),\hspace*{.8in}\forall k=1,...,p;\]
on the other hand, suppose partitions ${\cal P}\equiv \{C^1_1,...,C^{l_1}_1,...,C^1_p,...,C^{l_p}_p\}$ and ${\cal P}'\equiv \{C_1,...,C_p\}$ satisfy ${\cal P}'\in\mbox{argmax}_{\;{\cal P}''\in \mathscr U(N,{\cal P})}\tilde w(N,\vec v,{\cal P}'')$, then $\tilde w(N,\vec v,{\cal P}')\geq \tilde w(N,\vec v,{\cal P})$ if and only if
\[ \tilde w\left(C_k,\vec v|_{\mathscr C(C_k)},\{C_k\}\right)\geq \tilde w\left(C_k,\vec v|_{\mathscr C(C_k)},\{C^1_k,...,C^{l_k}_k\}\right),\hspace*{.8in}\forall k=1,...,p.\]
\end{fact}

\vspace*{-.1in}\noindent{\bf Proof: }For both the fission and fusion cases, the ``if'' part should be immediate without even using the hypothesized relationship between ${\cal P}$ and ${\cal P}'$; by~(\ref{worth-def}), 
\[ \tilde w\left(N,\vec v,\{C_1,...,C_p\}\right)=\sum_{k=1}^p\tilde w\left(C_k,\vec v|_{\mathscr C(C_k)},\{C_k\}\right),\] 
\[ \tilde w\left(N,\vec v,\{C^1_1,...,C^{l_1}_1,...,C^1_p,...,C^{l_p}_p\}\right)=\sum_{k=1}^p\tilde w\left(C_k,\vec v|_{\mathscr C(C_k)},\{C^1_k,...,C^{l_k}_k\}\right).\]

For the ``only if'' part, let us first tackle the fission case. Suppose $\tilde w(N,\vec v,{\cal P}')\geq \tilde w(N,\vec v,{\cal P})$ and yet for some $k=1,...,p$, 
\begin{equation}\label{contrast}
\tilde w\left(C_k,\vec v|_{\mathscr C(C_k)},\{C^1_k,...,C^{l_k}_k\}\right)<\tilde w\left(C_k,\vec v|_{\mathscr C(C_k)},\{C_k\}\right).
\end{equation}
This would be possible only when there is a $k'\neq k$ with $l_{k'}\geq 2$. Now construct a new partition ${\cal P}''\equiv {\cal P}'\hspace*{-.02in}\setminus\hspace*{-.03in}\{C^1_k,...,C^{l_k}_k\}\cup\{C_k\}$. Note ${\cal P}''$ is still a fission-down-to neighbor of ${\cal P}$. Using~(\ref{worth-def}), we may tell that
\[\tilde w(N,\vec v,{\cal P}'')=\tilde w(N,\vec v,{\cal P}')-\tilde w\left(C_k,\vec v|_{\mathscr C(C_k)},\{C^1_k,...,C^{l_k}_k\}\right)+\tilde w\left(C_k,\vec v|_{\mathscr C(C_k)},\{C_k\}\right),\]
which by~(\ref{contrast}), is strictly above $\tilde w(N,\vec v,{\cal P}')$. But this contradicts with the hypothesis that ${\cal P}'\in\mbox{argmax}_{\;{\cal P}''\in \mathscr I(N,{\cal P})}\tilde w(N,\vec v,{\cal P}'')$. Therefore, the opposite of~(\ref{contrast}) must be true for every $k=1,...,p$. The ``only if'' part of the fusion case can be similarly tackled. \qed

By Fact~\ref{p-motivation}, changes to a partition during SAM can occur one split/merger a time. However, it needs to be cautioned that a one-split fission or one-merger fusion need not be a one-step one. For instance, one coalition might split into three or more sub-coalitions and three or more coalitions might merge into one super-coalition.

%\bibliography{references}

\end{document}